\documentclass[nofootinbib,nobibnotes,groupaddress,preprintnumbers,aps,pra,twocolumn]{revtex4}
\usepackage{tipa,array}
\usepackage{color,multirow}
\usepackage{bm,epsfig,mathrsfs,amsmath,amssymb,graphicx,subfigure,overpic}
\usepackage{amsthm,amsmath,amssymb}
\usepackage{mathrsfs}
\usepackage{graphicx}
\usepackage{ulem}
\usepackage{dcolumn}
\usepackage{chngcntr}
\usepackage{upgreek}
\usepackage{float} 
\usepackage{booktabs}%
\usepackage{booktabs}%
\usepackage{amsmath}
\usepackage{bm}
\newcommand{\PreserveBackslash}[1]{\let\temp=\\#1\let\\=\temp}
\newcolumntype{C}[1]{>{\PreserveBackslash\centering}p{#1}}
\newcolumntype{R}[1]{>{\PreserveBackslash\raggedleft}p{#1}}
\newcolumntype{L}[1]{>{\PreserveBackslash\raggedright}p{#1}}
\usepackage[colorlinks,citecolor=blue,linkcolor=red]{hyperref}
\usepackage[toc,title,titletoc,header]{appendix}

\date{\today}
\begin{document}

\title{Local response theory of multiparameter sudden-quench quantum Otto engines}
\author{Dehua Liu$^{1,2}$}
\author{Yihang Cai$^{1}$}
\author{Biao Xiong$^1$}
\author{Jibing Liu$^1$}
\author{Chuanjia Shan$^1$}
\author{Jianhui Wang$^2$}
\email{liudehua@hbnu.edu.cn}
\email{cjshan@hbnu.edu.cn}
\email{wangjianhui@ncu.edu.cn}
\affiliation{ $^1\,$College of Physics and Electronic Science, Hubei Normal University, Huangshi 435002, China}
\affiliation{  $^2\,$Department of Physics, Nanchang University,
Nanchang 330031, China}
\begin{abstract} 
Multiparameter quantum Otto engines raise an optimization problem absent in single-parameter cycles: which small quench vector maximizes the quadratic approximation to the exact output work in a multidimensional control space? We develop a local response theory for sudden-quench Otto engines with quantum working media described by linearly controlled Hamiltonians. We show that the local quadratic work decomposes into a thermal-bias gain and a quadratic response cost, whose competition determines the quench vector that maximizes the local quadratic work at a prescribed quench amplitude.  We illustrate the theory in a minimal two-level working medium, where radial and transverse quenches play distinct
roles, and in a finite interacting XYZ spin chain, where the resulting  quench vector is determined by the balance between thermal-bias and response cost. Our results provide a practical local criterion for designing small multiparameter quenches in sudden-quench quantum Otto engines.


\end{abstract}

\maketitle
\date{\today}

\section{INTRODUCTION}

Quantum heat engines provide a paradigmatic setting for analyzing microscopic energy conversion between thermal reservoirs and extractable work \cite{Sco59,Ali79,Gev94,Fel00,Kos14,Goo16,Can24}. In a quantum Otto cycle, this conversion is implemented by alternating thermalization strokes with work strokes, during which external controls change the Hamiltonian of the working medium \cite{Qua07,Kos17,Aba12,Pet19,Ros16,Kla19,Bou21,Zha02}. In controlled quantum systems, such controls can be realized through external fields, coupling strengths, or interaction parameters \cite{Blo12,Bla12,Geo14,Mon21}. When these controls are varied in finite time, the work strokes generally drive the working medium away from quasistatic evolution \cite{Hol15,Ma24,Cak17,Aba19,Che19b,Klo19,Lee20,Sol20,Ins20,Sar21,Fei22,Xia23,McK25,Liu26,Ye22}. This makes the design of the work stroke central to finite-time engine operation.

A particularly simple limiting case of nonequilibrium work strokes is the sudden-quench protocol, in which the Hamiltonian is changed on a time scale shorter than the characteristic time scale of the state evolution.
In this limit, the density operator remains unchanged during each work stroke, and the work performed in the stroke is determined by the Hamiltonian change evaluated on the pre-quench state~\cite{Wat25,Wat26,Sil08,Cal11,Dor12,Arr19}. Thus, the choice of Hamiltonian modulation is not merely a matter of physical implementation, but an essential part of the engine design~\cite{Aba18,Cak19,Aba19,Ped23,Dan21, Don19}. 
Optimizing a sudden-quench Otto engine first requires specifying which available controls are used during the work strokes.
This observation naturally separates single-parameter cycles from multiparameter cycles.

Many quantum Otto engines are formulated with a single tunable control parameter~\cite{Kos17,Jar16,Che19,Kel20,Li18,Nau24,Saj25,Wat25}. In such settings, a quench modifies only one Hamiltonian component, so the work stroke is characterized solely by the quench amplitude. The control space is therefore one-dimensional, with no independent directional degree of freedom to optimize. This restriction is lifted in multiparameter engines, where several external fields, coupling strengths, or interaction parameters can be varied within the same cycle 
\cite{Koc23,Sim23,Wat25a,Wil24,Bou23, Wat26}. Multiparameter working media of this kind arise naturally in experimentally controllable quantum systems. Typical examples include few-level models with multiple control parameters~\cite{Wan12,Meh17}, spin systems with tunable fields and couplings~\cite{Azi14,Cho16,Saj25,Alt15}, and interacting many-body platforms with controllable interaction strengths and trapping parameters~\cite{Are24,Har20,Sol23,Sun18}.
With more than one control parameter, a quench acquires both a magnitude and a direction in a multidimensional control space.

Recent work on multiparameter sudden-quench Otto cycles has shown that simultaneous modulation of several Hamiltonian parameters can enhance thermodynamic performance relative to the corresponding single-parameter cycles \cite{Wat26}. For small quenches, that work further related
the work enhancement over the sum of the constituent single-parameter cycles to mixed derivatives of the high-energy Helmholtz free energy, or equivalently to
static susceptibilities. Its central question is therefore comparative: for specified low- and high-energy parameter
sets, how does simultaneous multiparameter control perform relative to the corresponding single-parameter cycles?
The present work addresses a complementary optimization problem. Once several control parameters are available, the quench vector itself becomes an optimization variable.
Near a given reference point, infinitely many small quench vectors are possible, and their directions and amplitudes
generally lead to different output work because the controlled operators are coupled through the equilibrium response of the working medium. The central question is
therefore not whether a given multiparameter stroke is beneficial relative to single-parameter operation, but how the direction and amplitude of a small quench should be
selected to maximize the local approximation to the total output work.

Local variations of Gibbs states can be characterized by equilibrium response and correlation matrices. Geometric approaches built from such local response structures have been used to quantify thermodynamic
length, irreversible cost, and performance bounds for prescribed finite-time or slow-driving protocols \cite{Siv12,Zul12,Def13,Bra20}. In those settings, the local response structure mainly evaluates the cost or geometry of a
given path. For a multiparameter sudden-quench Otto cycle, however, the central problem is different: one must select the direction and amplitude of a small quench from the available control space so as to
maximize the local approximation to the output work.

In this paper, we develop a local response theory for sudden quench Otto engines with quantum working media. We formulate the selection of a small multiparameter quench as an optimization problem for the local quadratic work. Starting from the exact sudden quench cycle work, we expand the hot and cold equilibrium expectation values around a common reference point in control space. The resulting local quadratic work separates into a thermal-bias gain and a quadratic response cost. In an energy eigenbasis of the reference Hamiltonian, the thermal-bias gain depends only on the diagonal matrix elements of the quench operator, whereas the quadratic response cost contains both diagonal and off diagonal matrix elements. Using equilibrium information at a single reference point, we diagonalize the response-cost matrix and decompose the local quadratic work into independent response eigenchannels. The balance between the thermal-bias projection and the response-cost eigenvalues in these channels determines the quench vector that maximizes the local quadratic work. We further prove that the quench vector obtained by maximizing the local quadratic work nearly maximizes the exact output work in the small amplitude limit.

We first illustrate the local quadratic work optimization using a two-level working medium, for which analytical expressions can be obtained. In this minimal model, the response eigenchannels reduce to radial and transverse directions in control space. The radial channel carries the thermal-bias gain and supports work
extraction, whereas the transverse channel has no thermal-bias projection
and contributes only a response cost. 
We then apply the same optimization procedure to a finite interacting XYZ spin chain as a multiparameter many body example.
 In this case, the quench vector that maximizes the local quadratic work is a collective combination of several response eigenchannels rather than a simple proportional modulation of the Hamiltonian parameters. These examples illustrate the role of the response eigenchannel decomposition. It reveals how thermal-bias gain competes with response cost in both analytically tractable models and interacting many body systems.

The remainder of this paper is organized as follows. In Sec.~\ref{sec1} we formulate the multiparameter sudden-quench quantum Otto engine and derive the exact work expression. In Sec.~\ref{sec2} we derive the local quadratic work, express the response-cost matrix in terms of BKM covariances, and obtain the unconstrained and constrained quench vectors that maximize the local quadratic work. In Sec.~\ref{sec3} we apply the theory to a two-level working medium and an interacting finite XYZ spin chain. Sec.~\ref{con} summarizes the main results and discusses possible extensions. 
Six appendices supplement the main text. Appendix \ref{appA} derives the
equilibrium-correlation form of the response matrix, Appendix \ref{appD} gives
the spectral representation of the thermal-bias gain and response
cost, and Appendix \ref{appC} presents the constrained local optimization.
Appendix \ref{appF} establishes the asymptotic near-optimality of the
constrained local optimizer for the exact output work. Appendix \ref{appB}
gives the detailed derivation of the response-cost eigenvalues for
the two-level model, and Appendix  \ref{appE} describes the benchmark obtained
by optimizing the exact output work at fixed amplitude in the XYZ spin-chain example.

\section{Multiparameter sudden-quench Otto cycle}
\label{sec1}
In this section, we formulate the multiparameter sudden-quench quantum Otto engine and derive the exact cycle work. We first introduce a linearly controlled working medium with several externally tunable parameters. We then specify the two sudden-quench work strokes and the two thermalization strokes of the Otto cycle. This gives the exact output work and efficiency, which serve as the starting point for the local theory developed below.

\subsection{Multiparameter working medium}
\label{sec11}
We consider a quantum working medium whose Hamiltonian depends linearly on \(M\) externally tunable parameters,
\begin{equation}
\hat H(\bm{\lambda})
=
\hat H_0+\sum_{\mu=1}^{M}\lambda_\mu \hat O_\mu .
\label{Hmulti}
\end{equation}

Here \(\bm{\lambda}=(\lambda_1,\lambda_2,\ldots,\lambda_M)^T\)
is the control-parameter vector. The operator \(\hat H_0\) denotes the
part of the Hamiltonian that is independent of the control parameters,
whereas \(\hat O_\mu\) is coupled to \(\lambda_\mu\). We assume that the
operators \(\hat O_\mu\) themselves do not depend on \(\bm{\lambda}\).

For compactness, we collect the controlled operators as
\(\hat{\bm O}=(\hat O_1,\hat O_2,\ldots,\hat O_M)^T\). For any state
\(\hat\rho\), we define $\langle\hat{\bm O}\rangle_{\hat\rho}
=
{\rm Tr}(\hat\rho\hat{\bm O}),$ where the trace is understood componentwise. When the working medium is thermalized at inverse temperature \(\beta\) for a fixed \(\bm{\lambda}\),
its equilibrium state is $\hat\rho_\beta(\bm{\lambda})=
e^{-\beta\hat H(\bm{\lambda})}/
Z_{\beta}({\bm\lambda}),$ where $Z_{\beta}(\bm\lambda)={\rm Tr}\,[e^{-\beta\hat H(\bm{\lambda})}]$ denotes the partition function. We set $\hbar=k_{B}=1$. The corresponding expectation value is 
\begin{equation}
    \langle \hat O_\mu\rangle_{\beta,\bm{\lambda}}={\rm Tr}[
\hat\rho_\beta(\bm{\lambda})\hat O_\mu].
\end{equation}

Although $\lambda_{\mu}$ couples directly to
$\hat{O}_{\mu}$, the expectation value
$\langle\hat{O}_{\mu}\rangle_{\beta,\boldsymbol{\lambda}}$
generally depends on the full control vector
$\boldsymbol{\lambda}$.
 Thus different control
channels can be coupled in equilibrium.

\subsection{Sudden-quench Otto cycle}
\label{sec12}
We consider a sudden quench from \(\bm{\lambda}_i\) to
\(\bm{\lambda}_f\). In the sudden-quench limit, the Hamiltonian is
changed externally while the state remains unchanged. The average work done on
the system is therefore
\begin{equation}
W_{i\to f}^{\rm sys}
=
{\rm Tr}\!\left[
\hat\rho_i
\left(
\hat H(\bm{\lambda}_f)-\hat H(\bm{\lambda}_i)
\right)
\right]=
(\bm{\lambda}_f-\bm{\lambda}_i)\cdot
\langle\hat{\bm O}\rangle_i.
\label{eq3}
\end{equation}

We now apply Eq. (\ref{eq3}) to a four-stroke Otto cycle consisting of two sudden quenches and two complete thermalization strokes. The working medium is first thermalized with the hot reservoir
at inverse temperature $\beta_h$, while the control vector is
held fixed at $\boldsymbol{\lambda}_h$, yielding
$\hat{\rho}_h
=\hat{\rho}_{\beta_h}(\boldsymbol{\lambda}_h)$.
It then undergoes a sudden quench from
$\boldsymbol{\lambda}_h$ to $\boldsymbol{\lambda}_c$,
followed by thermalization with the cold reservoir at inverse
temperature $\beta_c$($>\beta_h$), with the control vector held fixed at
$\boldsymbol{\lambda}_c$. The resulting equilibrium state is
$\hat{\rho}_c
=\hat{\rho}_{\beta_c}(\boldsymbol{\lambda}_c)$.
A second sudden quench then returns the Hamiltonian from
$\boldsymbol{\lambda}_c$ to
$\boldsymbol{\lambda}_h$.
The subsequent thermalization with the hot reservoir restores
the state $\hat{\rho}_h$ and closes the cycle. In the present idealized cycle, the two work strokes are
instantaneous quenches, while each thermalization stroke is
assumed to bring the working medium completely to the
corresponding Gibbs state.

The total average work done on the system during the two sudden quenches is \cite{Wat26}
\begin{align}
W_{\rm cyc}^{\rm sys}=
-(\bm{\lambda}_h-\bm{\lambda}_c)\cdot
\left[
\langle\hat{\bm O}\rangle_{\beta_h,\bm{\lambda}_h}
-\langle\hat{\bm O}\rangle_{\beta_c,\bm{\lambda}_c}
\right].
\label{wsys}
\end{align}

Hence the exact output work of
the sudden-quench Otto cycle is $W_{\rm out}=-W_{\rm cyc}^{\rm sys}$.
A positive value of \(W_{\rm out}\) corresponds to net work extraction.

The energy absorbed from the hot reservoir is the energy change
of the working medium during the hot thermalization stroke.
After the second sudden quench, the state remains $\hat{\rho}_c$, whereas the Hamiltonian has returned to
$\hat{H}({\bm \lambda}_h)$. The energy absorbed from the
hot reservoir is therefore
\begin{equation}
Q_h=\mathrm{Tr}
\left[\hat{H}(\bm\lambda_h)
(\hat{\rho}_h-\hat{\rho}_c)
\right].
\end{equation}

Similarly, the energy exchanged with the cold reservoir during the cold thermalization stroke is
$Q_c=\operatorname{Tr}
\left[
\hat{H}(\boldsymbol{\lambda}_c)
(\hat{\rho}_c-\hat{\rho}_h)
\right].$
The heat engine efficiency is defined as 
\begin{equation}
    \eta=\frac{W_{\rm out}}{Q_h},
\end{equation}
with the heat-engine regime specified by
$W_{\rm out}>0$, $Q_h>0$, and $Q_c<0$.

\section{LOCAL QUADRATIC WORK AND QUENCH OPTIMIZATION}
\label{sec2}
\subsection{Quadratic expansion of the exact output work}

\label{sec21}
We now analyze the local structure of the exact output work in the
multiparameter control space. We consider small Otto cycles around a
reference parameter vector \(\bm{\lambda}_0\), with
\begin{equation}
\bm{\lambda}_h
=
\bm{\lambda}_0+\frac{1}{2}\Delta\bm{\lambda},
\qquad
\bm{\lambda}_c
=
\bm{\lambda}_0-\frac{1}{2}\Delta\bm{\lambda}.
\end{equation}

The vector $\Delta\boldsymbol{\lambda}$ specifies both the direction and magnitude of the local quench. Throughout the main
text, we fix the units and normalization of the control coordinates
and define the quench amplitude by the equal-weight Euclidean norm
$\Vert\Delta\boldsymbol{\lambda}\Vert=
[\sum_{\mu}(\Delta\lambda_{\mu})^{2}]^{1/2}$.
 We refer to $\Delta\boldsymbol{\lambda}$ as the local quench vector, corresponding to the displacement of control parameters linking the two Hamiltonians employed during the sudden-quench work strokes.
Using Eq.~\eqref{wsys}, the exact sudden-quench work can be
written as
\(W_{\rm out}
=
\Delta\bm{\lambda}\cdot\bm b(\Delta\bm{\lambda})\),
where
$\bm b(\Delta\bm{\lambda})
=
\langle \hat{\bm O}\rangle_h
-
\langle \hat{\bm O}\rangle_c$.
At the reference point, we define the thermal-bias vector $\bm b_0
=
\langle \hat{\bm O}\rangle_{\beta_h,\bm{\lambda}_0}
-
\langle \hat{\bm O}\rangle_{\beta_c,\bm{\lambda}_0}.$
A first-order Taylor expansion of the two expectation values
around \(\bm{\lambda}_0\) gives
\begin{align}
\langle \hat{\bm O}\rangle_h
&=
\langle \hat{\bm O}\rangle_{\beta_h,\bm{\lambda}_0}
+
\frac{1}{2}
D^{(\beta_h)}
\Delta\bm{\lambda}
+
\mathcal{O}(\Vert\Delta\bm{\lambda}\Vert^2),
\nonumber\\
\langle \hat{\bm O}\rangle_c
&=
\langle \hat{\bm O}\rangle_{\beta_c,\bm{\lambda}_0}
-
\frac{1}{2}
D^{(\beta_c)}
\Delta\bm{\lambda}
+
\mathcal{O}(\Vert\Delta\bm{\lambda}\Vert^2),
\end{align}
where $D^{(\beta)}$ is the static response matrix evaluated at $\bm\lambda_0$, with elements
\begin{equation} 
\left(D^{(\beta_\alpha)}\right)_{\mu\nu}
=
\left.
\frac{\partial
\langle \hat O_\mu\rangle_{\beta_\alpha,\bm{\lambda}}}
{\partial \lambda_\nu}
\right|_{\bm{\lambda}=\bm{\lambda}_0},
\qquad
\alpha=h,c .
\label{response}
\end{equation}

It follows that $\bm b(\Delta\bm{\lambda})
=
\bm b_0
+
\frac{1}{2}
\left[
D^{(\beta_h)}+D^{(\beta_c)}
\right]
\Delta\bm{\lambda}
+
\mathcal{O}(\Vert\Delta\bm{\lambda}\Vert^2).$
Substituting this expression into the exact output work formula gives
$W_{\rm out}
=\Delta\bm{\lambda}\cdot\bm b_0
+\frac{1}{2}
\Delta\bm{\lambda}^{T}
[D^{(\beta_h)}+D^{(\beta_c)}]
\Delta\bm{\lambda}
+
\mathcal{O}(\Vert\Delta\bm{\lambda}\Vert^3).$
Introducing $ K=-
[D^{(\beta_h)}+D^{(\beta_c)}]/2,$
 the local quadratic work can be written as 
\begin{equation}
W_{\rm loc}
=
\Delta\bm{\lambda}\cdot\bm b_0
-
\Delta\bm{\lambda}^{T} K\Delta\bm{\lambda}.
\label{localwork}
\end{equation}

Thus,
\(W_{\rm out}
=W_{\rm loc}+
\mathcal{O}(\Vert\Delta\bm{\lambda}\Vert^3)\).
Equation~\eqref{localwork} separates the local quadratic work into two contributions:
a linear gain set by the thermal-bias vector at the reference point, and a quadratic term determined by how the expectation values respond to changes in the control
parameters.


\subsection{Bogoliubov--Kubo--Mori covariance form of $K$}
We now express the quadratic coefficient \(K\) in terms of equilibrium
correlations. For the linearly controlled Hamiltonian in Eq.~(\ref{Hmulti}), the static response matrix satisfies
\begin{equation}
D^{(\beta)}
=-\beta C^{(\beta)} .
\end{equation}

Here $C^{(\beta)}$ denotes the Bogoliubov–Kubo–Mori (BKM) covariance matrix
\cite{Per24,Lac25, Bet25}, whose elements are
$(C^{(\beta)})_{\mu\nu}
=\int_0^1 ds\,
\mathrm{Tr}[
\rho_\beta^s\delta\hat O_\mu
\rho_\beta^{1-s}\delta\hat O_\nu
]$,
with
$\delta\hat O_\mu
=\hat O_\mu-\langle\hat O_\mu\rangle_{\beta,\boldsymbol{\lambda}}\hat{I}$.

Using the definition \(K=-[D^{(\beta_h)}+D^{(\beta_c)}]/2\), the matrix \(K\) can be written as
\begin{equation}
K
=
\frac{1}{2}
\left[
\beta_h C^{(\beta_h)}
+
\beta_c C^{(\beta_c)}
\right],
\label{Kcorrelation}
\end{equation}
where the two BKM covariance matrices are evaluated at
\((\beta_h,\bm\lambda_0)\) and \((\beta_c,\bm\lambda_0)\), respectively.

For Hermitian controlled operators, \(C^{(\beta)}\) is positive
semidefinite, and hence so is \(K\) for \(\beta_h,\beta_c>0\). Therefore,
\(\Delta\boldsymbol{\lambda}^{T}K\Delta\boldsymbol{\lambda}\ge0\).  Since this quantity enters Eq.~(\ref{localwork}) with a minus sign, it represents the quadratic response cost of the quench. We therefore refer to $K$ as the response-cost matrix. A detailed derivation of the response relation and the positive
semidefiniteness of \(K\) is given in Appendix~\ref{appA}.



\subsection{Spectral interpretation of the local quadratic work}
\label{sec24}

The local quadratic work in Eq.~(\ref{localwork}) consists of a linear
thermal-bias gain $\Delta\bm \lambda\cdot\bm b_0$ and a quadratic response cost $\Delta\bm\lambda^{T} K\Delta\bm\lambda$. To make the microscopic content of these two terms explicit, we express
them in the energy eigenbasis of the reference Hamiltonian
$\hat H_{\rm r}|n\rangle=E_n|n\rangle,$ where $\hat{H}_r\equiv\hat{H}(\bm\lambda_0)$.
 The reference Gibbs state on branch \(\alpha=h,c\) is
$\hat\rho_{\alpha}
\equiv\hat\rho_{\beta_\alpha}(\boldsymbol{\lambda_0})=
\sum_n p_n^{(\alpha)}|n\rangle\langle n|,$ where 
$p_n^{(\alpha)}
=e^{-\beta_\alpha E_n}/Z_\alpha.$

For the local quench vector
\(\Delta\boldsymbol{\lambda}\),
the corresponding Hamiltonian change is
$\hat X\equiv\hat H(\boldsymbol{\lambda}_h)-\hat H(\boldsymbol{\lambda}_c)=\sum_\mu \Delta\lambda_\mu \hat O_\mu.$
We also define its centered form on branch \(\alpha\) as
$\delta\hat X_\alpha
=
\hat X-\langle\hat X\rangle_{\beta_\alpha}\hat I.$
Here $X_{mn}\equiv\langle m|\hat X|n\rangle$ and
\((\delta X_\alpha)_{mn}\equiv\langle m|\delta\hat X_\alpha|n\rangle\).
 The thermal-bias gain can be written as
\begin{equation}
\Delta\boldsymbol{\lambda}\cdot\boldsymbol b_0=\sum_n
\left(
p_n^{(h)}-p_n^{(c)}
\right)\hat{X}_{nn},
\label{dlb}
\end{equation}

Here, the linear thermal-bias gain is determined only by the energy-diagonal part of the quench operator.

By contrast, the quadratic response cost contains both diagonal matrix elements and matrix elements that are off diagonal in the energy eigenbasis,
\begin{equation}
\Delta\boldsymbol{\lambda}^{T}K\Delta\boldsymbol{\lambda}
=\frac12
\sum_{\alpha=h,c}
\beta_\alpha
\sum_{m,n}
L\!\left(
p_m^{(\alpha)},p_n^{(\alpha)}
\right)
\left|
(\delta X_\alpha)_{mn}
\right|^2,
\label{cost1}
\end{equation}
where
\begin{equation}
L(x,y)
=
\int_0^1 ds\,x^s y^{1-s}
=
\begin{cases}
\dfrac{x-y}{\ln x-\ln y}, & x\neq y,\\[6pt]
x, & x=y .
\end{cases}
\label{L1}
\end{equation}

Equation~(\ref{cost1}) shows that the quadratic response cost contains matrix elements within the same energy eigenspace as well as those connecting distinct energy eigenspaces. For degenerate spectra, this distinction is understood with respect to energy eigenspaces rather than individual eigenstates (see Appendix \ref{appD} for details).

\subsection{Local optimization of the quench vector}
\label{sec23}

We now determine the quench vector that maximizes the local quadratic work in the small-quench regime. 
In the original control basis, the response-cost matrix $K$ generally contains off-diagonal elements that couple different control parameters. We therefore diagonalize $K$ to identify independent response channels. Since $K$ is real symmetric and positive semidefinite, it admits an orthonormal eigenbasis,
\begin{equation}
    K \boldsymbol{v}_k = \kappa_k \boldsymbol{v}_k,
\end{equation}
where $\boldsymbol{v}_k^\mathrm{T} \boldsymbol{v}_l = \delta_{kl}$ and $\kappa_k \ge 0$ is response-cost eigenvalue. We expand the local quench vector in this eigenbasis as $\Delta\bm\lambda = \sum_k \xi_k v_k$, where $\xi_k$ is the amplitude along the $k$th response eigenchannel. In this basis, the quadratic response cost becomes $\Delta\bm\lambda^T K \Delta\bm\lambda = \sum_k \kappa_k \xi_k^2$. The thermal-bias gain is obtained by projecting $\bm b_0$ onto the same eigenbasis, with $b_k = v_k^T \bm b_0$, so that $\Delta\bm\lambda \cdot \bm b_0 = \sum_k \xi_k b_k$. 
Substituting these expansions into Eq.~(\ref{localwork}) gives the eigenchannel form of the local quadratic work,
\begin{equation}
W_{\rm loc}
=
\sum_{\kappa_k>0}
\left(b_k
\xi_k -\kappa_k\xi_k^2
\right).
\label{work}
\end{equation}

Equation~(\ref{work}) decomposes the local quadratic work  into independent response eigenchannels. In the $k$th channel, $b_k$ measures the work gain generated by the thermal-bias, while $\kappa_k$ quantifies
the corresponding response cost. The contribution of each response channel is determined jointly by its thermal-bias projection $b_k$ and response cost eigenvalue $\kappa_k$.

Since $W_{\rm loc}$ is valid only for small quenches, we optimize it within the region $B_\epsilon=\{\Delta\boldsymbol{\lambda}:
\Vert\Delta\boldsymbol{\lambda}\Vert\leq\epsilon\}$. The bound $\epsilon$ is chosen independently and is not determined by the local theory. In the numerical examples, we test the validity of this bound by comparing
$W_{\rm loc}$ with $W_{\rm out}$. More generally, different control parameters may be assigned different weights through the constraint $\Delta\boldsymbol{\lambda}^{T}G\Delta\boldsymbol{\lambda}
\leq\epsilon^2$.
The equal-weight Euclidean norm used here corresponds to $G=I$.

In the response eigenbasis, the local quadratic work contains no coupling between different channel amplitudes. Therefore, before imposing the amplitude constraint $\Vert\Delta\boldsymbol{\lambda}\Vert \leq \epsilon$, the local quadratic work can be maximized independently with respect to each channel amplitude $\xi_k$. For each channel with $\kappa_k > 0$, maximizing $b_k \xi_k - \kappa_k \xi_k^2$ gives
$\xi_{k,{\text{unc}}} =b_k/2\kappa_k.$
This yields the unconstrained quench vector,
\begin{equation}
\Delta\bm{\lambda}_{\rm unc}
=\frac{1}{2}
\sum_{\kappa_k>0}
\frac{b_k}{\kappa_k}\bm v_k .
\label{unc}
\end{equation}


If $\Vert\Delta\boldsymbol{\lambda}_{\text{unc}}\Vert \leq \epsilon$, then $\Delta\boldsymbol{\lambda}_{\text{unc}}$ lies in $B_\epsilon$ and also maximizes $W_{\text{loc}}$ over this admissible region. In this case, the maximum local quadratic work is
\begin{equation}
W_{\rm loc,\rm unc}^{\max}
=\sum_{\kappa_k>0}\frac{b_k^2}{4\kappa_k}.
\label{wmax}
\end{equation}


When $\Vert\Delta\boldsymbol{\lambda}_{\text{unc}}\Vert > \epsilon$, $\Delta\boldsymbol{\lambda}_{\text{unc}}$ lies outside the admissible region $B_\epsilon$. The quench vector that maximizes $W_{\text{loc}}$ over $B_\epsilon$ must then lie on the boundary $\Vert\Delta\boldsymbol{\lambda}\Vert = \epsilon$.  
Introducing a Lagrange multiplier $\zeta>0$ to impose this boundary condition, the optimal quench vector is
\begin{equation}
\Delta\bm{\lambda}^{\epsilon}_{\rm opt}
=\frac{1}{2}
\sum_{\kappa_k>0}
\frac{b_k}{\kappa_k+\zeta}\bm v_k ,
\label{con1}
\end{equation}
where $\zeta$ is determined by
$\sum_{\kappa_k>0}b_k^2/[4(\kappa_k+\zeta)^2]=\epsilon^2$. 

Substituting Eq.~(\ref{con1}) into Eq.~(\ref{localwork}) gives the
 maximum local quadratic work,
\begin{equation}
W_{\rm loc,\epsilon}^{\max}
=
\frac{1}{4}
\sum_{\kappa_k>0}
\frac{b_k^2(\kappa_k+2\zeta)}
{(\kappa_k+\zeta)^2}.
\label{wmaxcon}
\end{equation}

For an inactive constraint, $\zeta=0$, and this expression reduces to
Eq.~(\ref{wmax}). Therefore, each response channel can be optimized independently without an amplitude constraint. When the total quench amplitude is constrained, the optimal channel amplitudes must be determined together through the common Lagrange multiplier $\zeta$.
Complete derivations
of Eqs. (\ref{unc})–(\ref{wmaxcon})  are presented in Appendix \ref{appC}.

Suppose that the Taylor remainder satisfies
$\left|W_{\rm out}(\Delta\boldsymbol{\lambda})-W_{\rm loc}(\Delta\boldsymbol{\lambda})
\right|
\leq
C\Vert\Delta\boldsymbol{\lambda}\Vert^3$ throughout the admissible region $B_{\epsilon}$, where $C>0$ is a constant independent of $\Delta\bm\lambda$ within this region.
Let $\Delta\boldsymbol{\lambda}^{\epsilon}_{\rm ext}$ denote a maximizer of the exact
output work over $B_\epsilon$, then
\begin{equation}
0\leq W_{\rm out}
\left(\Delta\boldsymbol{\lambda}^{\epsilon}_{\rm ext}
\right)-W_{\rm out}
\left(
\Delta\boldsymbol{\lambda}^{\epsilon}_{\rm opt}
\right)
\leq
2C\epsilon^3.
\label{wml}
\end{equation}

Thus, the exact output work evaluated at the quench vector that maximizes $W_{\text{loc}}$ over $B_\epsilon$ differs from the maximum of $W_{\text{out}}$ over the same region by at most a term of order $\epsilon^3$ (see  Appendix~\ref{appF} for details).

Under the present assumptions of finite positive inverse temperatures and a finite-dimensional working medium, the Gibbs states are full rank. Since the BKM variance of a Hermitian operator in a full-rank state vanishes only when the operator is proportional to the identity \cite{Per24}, a zero response cost eigenvalue $\kappa_k = 0$ implies that the corresponding controlled-operator combination is proportional to the identity. It then follows that $b_k = 0$. Such null channels do not contribute to $W_\mathrm{loc}$ and may be omitted from the optimization.

To assess the effect of the local quadratic work approximation on the efficiency, we define
\begin{equation}
    \eta_{\mathrm{dig}} = \frac{W_{\mathrm{loc}}}{Q_h},
\end{equation}
where $Q_h$ is evaluated from the exact cycle. This quantity is not a separate approximation to the thermodynamic efficiency, because only the output work is approximated. It is introduced solely to measure the efficiency error caused by replacing $W_{\mathrm{out}}$ with $W_{\mathrm{loc}}$.

\section{Applications to representative working media}
\label{sec3}
We now apply the local quadratic work formalism to two representative working media. The first is a two-level system with two independent control parameters. This model is analytically tractable, and its response eigenchannels can be obtained explicitly. The second is an interacting XYZ spin chain with several control parameters. It extends the analysis to a higher-dimensional control space and provides a many-body realization of the sudden-quench Otto cycle. These two examples demonstrate that the same formalism applies to both a simple two-level engine and an interacting many-body system.

\subsection{Minimal two-level working medium}
\label{sec31}
We first illustrate the local eigenchannel construction using a minimal
two-level working medium with two independent control parameters. We
consider the Hamiltonian
\begin{equation}
\hat H(\bm{\lambda})
=\lambda_x \hat\sigma_x
+\lambda_z \hat\sigma_z ,
\qquad
\bm{\lambda}=(\lambda_x,\lambda_z)^T .
\end{equation}

The two control parameters determine both the level splitting and the
direction of the effective field in parameter space. The controlled-operator
vector is \(\hat{\bm O}=(\hat\sigma_x,\hat\sigma_z)^T\). Defining
\(r=|\bm\lambda|\) and \(\bm n=\bm\lambda/r\), the equilibrium expectation
value takes the compact form
\begin{equation}
\langle \hat{\bm O}\rangle_{\beta,\bm{\lambda}}
=
-\tanh(\beta r)\bm n .
\label{twoexp}
\end{equation}

At the reference point \(\bm{\lambda}_0=r_0\bm n_0\), the
thermal-bias vector becomes 
\begin{align}
    \bm b_0
=&\langle \hat{\bm O}\rangle_{\beta_h,\bm{\lambda_0}}-\langle \hat{\bm O}\rangle_{\beta_c,\bm{\lambda_0}}
\nonumber\\
=&
\left[\tanh(\beta_c r_0)-\tanh(\beta_h r_0)
\right]\bm n_0.
\label{b0}
\end{align}

This shows that the thermal-bias vector is purely radial in parameter
space. To obtain the static response matrix, we differentiate
Eq.~(\ref{twoexp}) with respect to the control parameters and
evaluate the result at \(\bm\lambda=\bm\lambda_0\). This gives
\begin{equation}
D^{(\beta)}
=-\beta\,{\rm sech}^2(\beta r_0)\,
\bm n_0\bm n_0^T
-\frac{\tanh(\beta r_0)}{r_0}
\left(\mathbb{I}-\bm n_0\bm n_0^T
\right).
\label{tworesponse}
\end{equation}

Equation~\eqref{tworesponse} has a radial--transverse projector structure. The first term acts along the reference direction $\bm n_0$, whereas the second term acts in the transverse subspace. 
It is therefore natural to introduce the radial direction \(\bm v_\parallel=\bm n_0\) and a unit transverse direction \(\bm v_\perp\) satisfying \(\bm v_\perp^T\bm n_0=0\).  Since \(D^{(\beta_h)}\) and
\(D^{(\beta_c)}\) share this projector structure, the matrix \(K\) is
diagonal in the same basis: 
$K\bm v_\parallel=\kappa_\parallel\bm v_\parallel$ and $
K\bm v_\perp=\kappa_\perp\bm v_\perp.$ The corresponding eigenvalues are
\begin{equation}
\kappa_{\parallel}
=
\frac{1}{2}
\left[
\beta_h{\rm sech}^2(\beta_h r_0)
+
\beta_c{\rm sech}^2(\beta_c r_0)
\right],
\label{kp}
\end{equation}
and
\begin{equation}
\kappa_{\perp}
=
\frac{1}{2r_0}
\left[
\tanh(\beta_h r_0)
+
\tanh(\beta_c r_0)
\right].
\label{kc}
\end{equation}

These eigenvalues determine the quadratic response costs associated with quenches along the radial and transverse response eigenchannels, respectively.
Detailed derivations of Eqs.~\eqref{tworesponse}--\eqref{kc} are
provided in Appendix~\ref{appB}.

Since the thermal-bias vector in Eq.~(\ref{b0}) is parallel to
\(\bm n_0\), its projections onto the radial and transverse response
eigenchannels are
\begin{equation}
b_\parallel=\bm v_\parallel^T\bm b_0
=\tanh(\beta_c r_0)-\tanh(\beta_h r_0),
\end{equation}
and
\begin{equation}
b_\perp=\bm v_\perp^T\bm b_0=0 .
\end{equation}

An arbitrary local quench vector can be written as
\(\Delta\bm\lambda=\xi_\parallel\bm v_\parallel+\xi_\perp\bm v_\perp\). 
Substitution into Eq.~\eqref{work} gives
\begin{equation}
W_{\rm loc}
=\xi_\parallel b_\parallel
-\kappa_\parallel \xi_\parallel^2
-\kappa_\perp \xi_\perp^2 .
\label{workt}
\end{equation}

\begin{figure}[t]
\begin{overpic}[width=3.3cm]{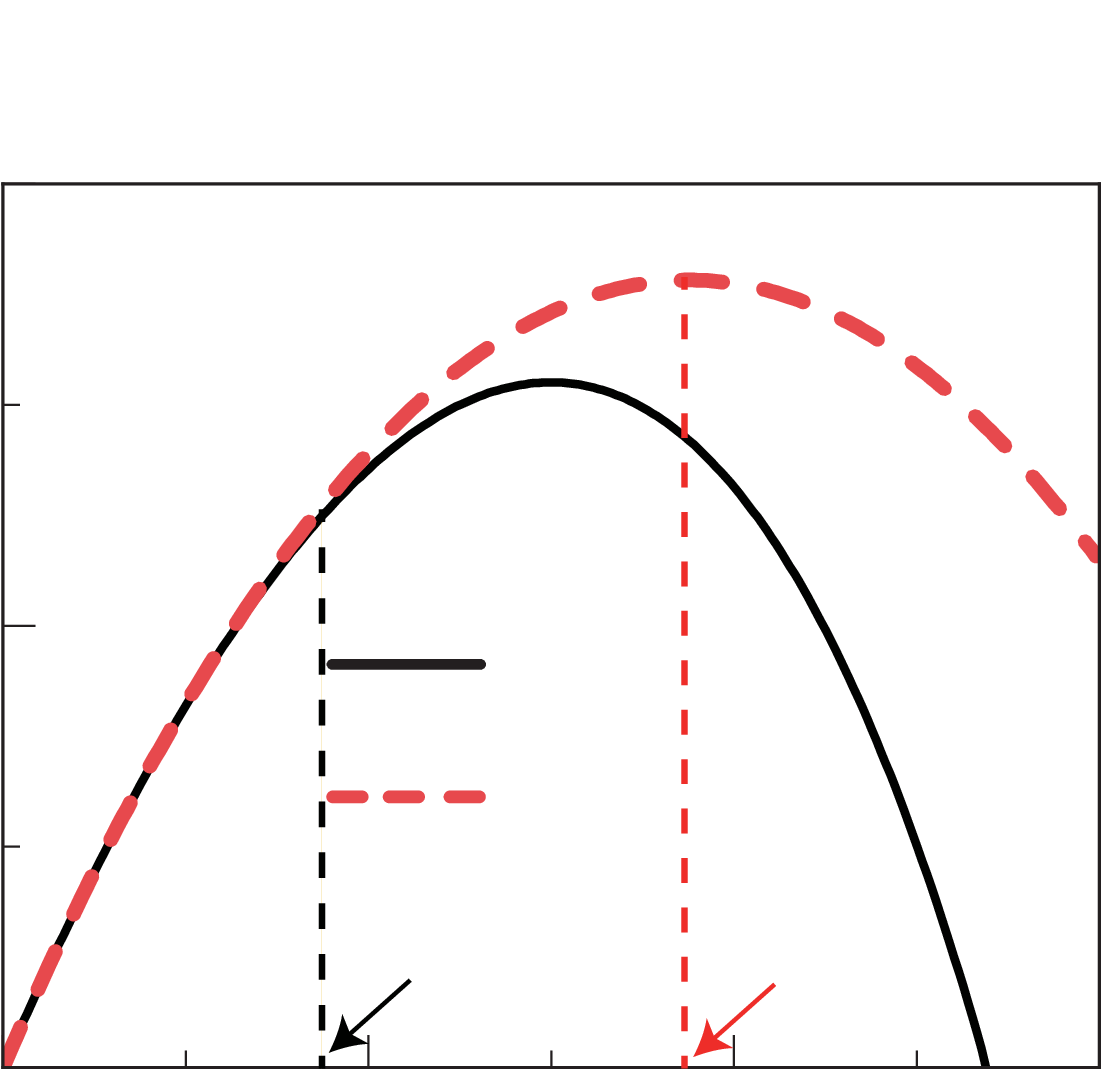}
\put(-22,88){\scalebox{0.9}{(a)}}
\put(-1,-7){\scalebox{0.9}{0}}
\put(28,-7){\scalebox{0.9}{0.4}}
\put(61.5,-7){\scalebox{0.9}{0.8}}
\put(92,-7){\scalebox{0.9}{1.2}}
\put(-5,-2){\scalebox{0.9}{0}}
\put(-16.5,37){\scalebox{0.9}{0.09}}
\put(-16.5,78){\scalebox{0.9}{0.18}}
\put(46,34){\scalebox{0.8}{$W_{\rm out}$}}
\put(46,22){\scalebox{0.8}{$W_{\rm loc}$}}
\put(45,-16){\scalebox{1.1}{$\xi_{\parallel}$}}
\put(36,7){\scalebox{0.8}{0.35}}
\put(69,7){\scalebox{0.8}{0.75}}
\end{overpic}
\hspace{0.3in}
\begin{overpic}[width=3.3cm]{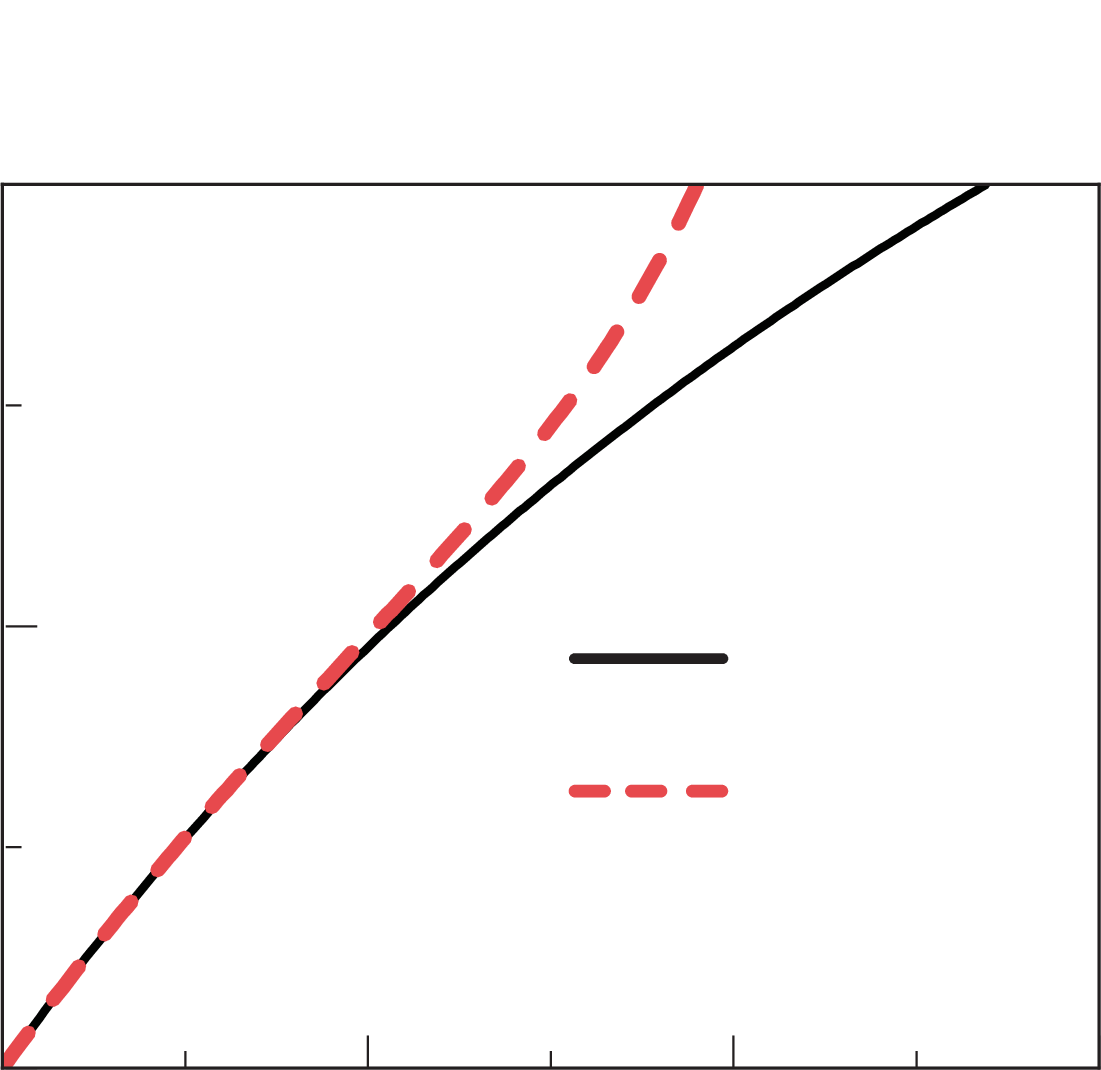}
\put(-22,88){\scalebox{0.9}{(b)}}
\put(-1,-7){\scalebox{0.9}{0}}
\put(28,-7){\scalebox{0.9}{0.4}}
\put(61.5,-7){\scalebox{0.9}{0.8}}
\put(92,-7){\scalebox{0.9}{1.2}}
\put(-5,-2){\scalebox{0.9}{0}}
\put(-12.5,38){\scalebox{0.9}{0.3}}
\put(-12.5,78){\scalebox{0.9}{0.6}}
\put(69,24){\scalebox{0.9}{$\eta_{\rm dig}$}}
\put(69,36){\scalebox{0.9}{$\eta$}}
\put(45,-16){\scalebox{1.1}{$\xi_{\parallel}$}}
\end{overpic}
\hspace{0.4in}
\caption{(a) Exact output work $W_{\rm out}$ and local quadratic work $W_{\rm loc}$, and
(b) exact efficiency $\eta$ and the diagnostic quantity $\eta_{\rm dig}$, as functions of the radial quench amplitude $\xi_\parallel$ along the path $\Delta\bm\lambda(\xi_{\parallel})$. The black solid and red dashed lines denote the exact and local results, respectively. 
The black and red vertical dashed lines mark the pathwise
$2\%$-deviation benchmark
$\xi_\parallel\simeq0.35$ and the unconstrained optimal radial
amplitude $b_\parallel/(2\kappa_\parallel)\simeq0.75$,
respectively.
The other parameters are
$\bm{\lambda}_0=(0.6,0.8)$,
$\beta_h=0.6$, and $\beta_c=2.0$.}
\label{a1}
\end{figure}

This result is consistent with the spectral interpretation in
Sec.~\ref{sec24}. At the reference point, the radial-channel operator
\(\boldsymbol v_\parallel\cdot\hat{\boldsymbol{O}}\)
is proportional to \(\hat H(\boldsymbol{\lambda}_0)\) and therefore
commutes with it, whereas the transverse-channel operator
$\boldsymbol v_\perp\cdot\hat{\boldsymbol O}$ is off diagonal in
the energy eigenbasis of $\hat H(\boldsymbol{\lambda}_0)$. Therefore, the transverse component
produces no linear thermal-bias gain and contributes only a  noncommuting response cost, represented by
\(-\kappa_\perp\xi_\perp^2\).
Using Eq.~\eqref{unc} together with $b_\perp = 0$, we find that the transverse amplitude vanishes. Therefore, for the optimization without the condition $\Vert\Delta\boldsymbol{\lambda}\Vert \leq \epsilon$, the unconstrained quench vector that maximizes $W_{\text{loc}}$ is purely radial, and the corresponding maximum local quadratic work is
\begin{equation}
\Delta\bm{\lambda}_{\rm unc}
=\frac{b_{\parallel}}{2\kappa_{\parallel}}\bm v_{\parallel},
\qquad
W_{\rm loc}^{\max}
=
\frac{b_{\parallel}^{2}}{4\kappa_{\parallel}}.
\label{twoun}
\end{equation}

\begin{figure}[t]
\begin{overpic}[width=3.3cm]{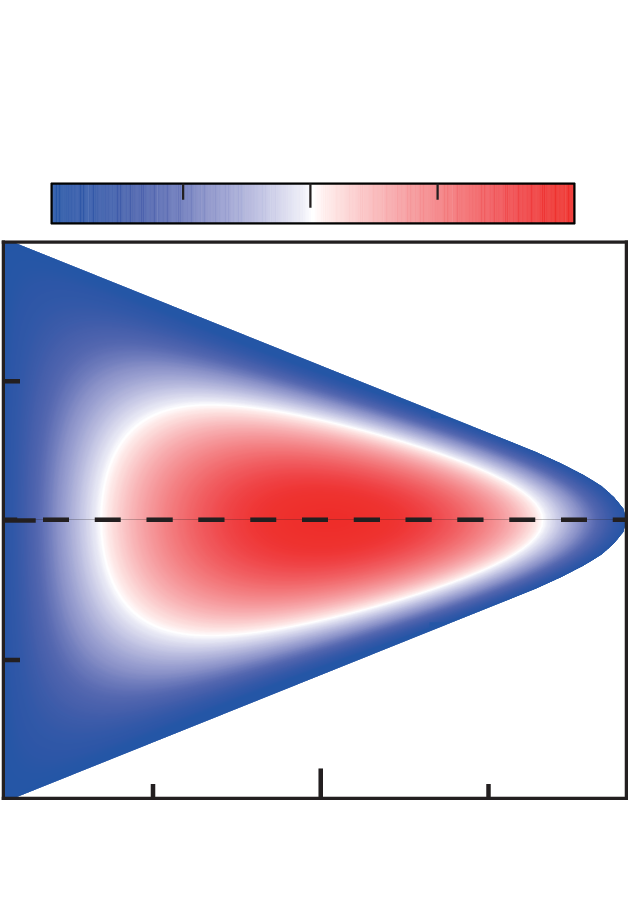}
\put(-12,88){\scalebox{0.9}{(a)}}
\put(-1,7){\scalebox{0.9}{0}}
\put(28,7){\scalebox{0.9}{0.503}}
\put(64,7){\scalebox{0.9}{1.06}}
\put(-15,11){\scalebox{0.9}{$-\pi/2$}}
\put(-5,41){\scalebox{0.9}{$0$}}
\put(-11,71){\scalebox{0.9}{$\pi/2$}}
\put(5,81){\scalebox{0.8}{0}}
\put(30,81){\scalebox{0.7}{0.07}}
\put(58,81){\scalebox{0.7}{0.14}}
\put(29,87){\scalebox{0.8}{$W_{\rm out}$}}
\put(34,0){\scalebox{1.2}{$\xi$}}
\put(-15,40){\rotatebox{90}{\scalebox{1.1}{$\theta$}}}
\end{overpic}
\hspace{0.4in}
\begin{overpic}[width=3.3cm]{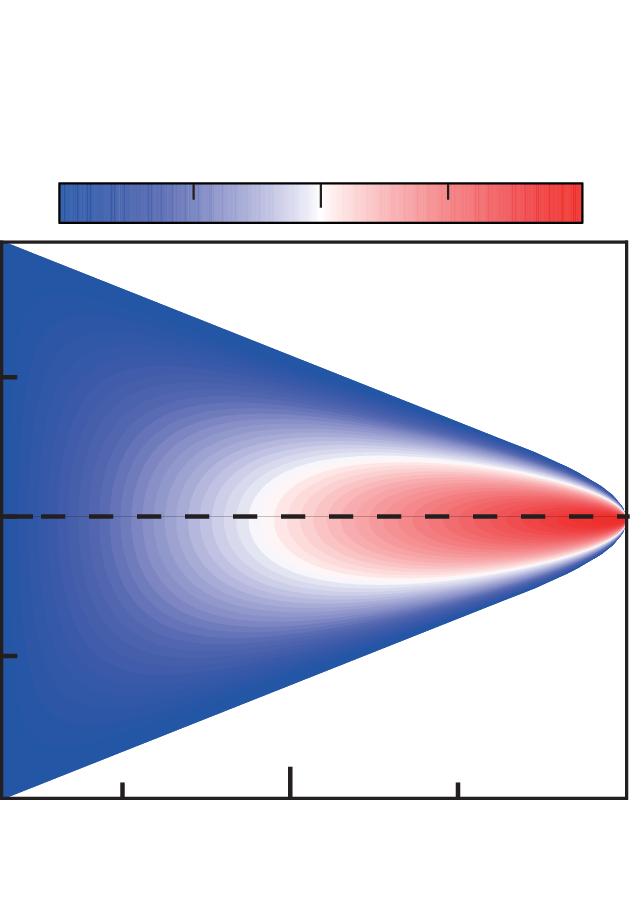}
\put(-12,88){\scalebox{0.9}{(b)}}
\put(5,81){\scalebox{0.8}{0}}
\put(30,81){\scalebox{0.7}{0.349}}
\put(58,81){\scalebox{0.7}{0.698}}
\put(-1,7){\scalebox{0.9}{0}}
\put(24,7){\scalebox{0.9}{0.503}}
\put(64,7){\scalebox{0.9}{1.06}}
\put(-15,11){\scalebox{0.9}{$-\pi/2$}}
\put(-5,41){\scalebox{0.9}{$0$}}
\put(-11,71){\scalebox{0.9}{$\pi/2$}}
\put(32,87){\scalebox{1.05}{$\eta$}}
\put(-15,40){\rotatebox{90}{\scalebox{1.1}{$\theta$}}}
\put(30,0){\scalebox{1.2}{$\xi$}}
\end{overpic}
\caption{Performance contour plots for the sudden-quench Otto cycle with the two-level working medium. The exact output work (a) and exact efficiency (b) are displayed in the quench-coordinate plane
parameterized by the amplitude $\xi$ and angle $\theta$.
 The other parameters are the  same as in Fig. \ref{a1}.}
\label{a2}
\end{figure}

The exact output work for the two-level working medium is
obtained by substituting Eq.~(\ref{twoexp}) into the exact work
expression. This gives
\begin{equation}
W_{\rm out}
=\Delta\bm{\lambda}\cdot
\left[\tanh(\beta_c r_c)\bm n_c
-\tanh(\beta_h r_h)\bm n_h
\right],
\label{WexactTLS}
\end{equation}
where \(r_\alpha=|\bm{\lambda}_\alpha|\) and
\(\bm n_\alpha=\bm{\lambda}_\alpha/r_\alpha\) for
\(\alpha=h,c\).

To examine how accurately $W_{\text{loc}}$ approximates the exact output work $W_{\text{out}}$ along the radial direction, we introduce the radial quench path $\Delta\bm\lambda(\xi_\parallel)=\xi_\parallel\bm v_\parallel$.
Figure~\ref{a1}(a) compares the exact output work $W_{\text{out}}$ with the local quadratic work $W_{\text{loc}}$ as functions of $\xi_\parallel$ along this path. Their agreement is quantified by the relative deviation
$\delta W = |W_{\text{out}} - W_{\text{loc}}|/|W_{\text{out}}|.$
For the present parameters, $\delta W$ remains below $2\%$ for $\xi_\parallel \lesssim 0.35$. This $2\%$ threshold is used only as a numerical criterion for the agreement between $W_{\text{loc}}$ and $W_{\text{out}}$ along the radial quench path and has no special physical significance. We therefore set $\epsilon = 0.35$ and restrict the subsequent optimization to the region $\Vert\Delta\boldsymbol{\lambda}\Vert \leq \epsilon$. As shown in Fig.~\ref{a1}(a), maximizing $W_{\text{loc}}$ with respect to the radial amplitude gives $\xi_{\parallel,\text{unc}} \simeq 0.75$, which exceeds the selected upper bound. The maximum of $W_{\text{loc}}$ is therefore attained at $\Vert\Delta\boldsymbol{\lambda}\Vert = \epsilon$. Since $b_\perp = 0$, Eq.~\eqref{con1} gives optimal quench vector
$\Delta\boldsymbol{\lambda}_{\text{opt}}^\epsilon = \epsilon \boldsymbol{v}_\parallel,$
with the corresponding maximum local quadratic work
$W_{\text{loc},\epsilon}^{\text{max}} = b_\parallel \epsilon - \kappa_\parallel \epsilon^2.$
For $\xi_\parallel > 0.35$, the difference between $W_{\text{loc}}$ and $W_{\text{out}}$ increases because the higher-order corrections neglected in $W_{\text{loc}}$ become more significant.

\begin{table}
\caption{Eigenchannel-resolved quantities for the XYZ spin chain. For each eigenchannel $k$, optimizing the contribution of that eigenchannel to $W_{\rm loc}$ gives the amplitude $b_k/(2\kappa_k)$ and the maximum contribution $W_{k,\rm loc}^{\rm max} = b_k^2/(4\kappa_k)$. The quantity $q_k^{\text{nc}} = \kappa_k^{\text{nc}}/\kappa_k$ is the noncommuting fraction of $\kappa_k$. The parameters are $N=5$, $\beta_h=0.6$,
$\beta_c=2$, and
$\boldsymbol{\lambda}_0=(0.1,0.8,0.4,1.6)$.
}
\label{tab}
\begin{ruledtabular}
\begin{tabular}{ccccccc}
Mode & $\kappa_k$ & $b_k$&  $b_k/(2\kappa_k)$  & $W_{k,\rm loc}^{\max}$ & $q_k^{\rm nc}$ \\
\hline
1 & 0.1218 & 0.1542& 0.6330   & 0.0488 & 0.3993 \\
2 & 0.8064 & 0.7457&  0.4623  & 0.1724 & 0.2440 \\
3 & 2.2280 & -0.0500&  -0.0112  & 0.0003 & 0.9589  \\
4 & 9.6779 & 1.9922& 0.1029  &  0.1025  & 0.0643 \\ 
\end{tabular}
\end{ruledtabular}
\end{table}

Along the same radial quench path, Fig.~\ref{a1}(b) compares the exact efficiency $\eta$ with the diagnostic quantity $\eta_{\text{dig}}$. The same exact heat input $Q_h$ is used to evaluate both $\eta$ and $\eta_{\text{dig}}$. Therefore, any difference between $\eta$ and $\eta_{\text{dig}}$ arises solely from using $W_{\text{loc}}$ instead of $W_{\text{out}}$ in $\eta_{\text{dig}}$. For $\xi_\parallel \lesssim 0.35$, $\eta_{\text{dig}}$ remains close to $\eta$, indicating that this replacement introduces only a small error in the efficiency. As $\xi_\parallel$ increases, the higher-order corrections neglected in $W_{\text{loc}}$ become more important, and the difference between $\eta_{\text{dig}}$ and $\eta$ increases accordingly. Hence, this difference originates from the approximation of $W_{\text{out}}$, rather than from any approximation of $Q_h$.

To show how the exact output work varies with the radial and transverse quench components, we plot $W_{\mathrm{out}}$ and   $\eta$ in the radial--transverse plane. The quench is parameterized as $\Delta\boldsymbol{\lambda} = \xi \left(\cos\theta \boldsymbol{v}_{\parallel} + \sin\theta \boldsymbol{v}_{\perp}\right)$, where $\xi$ is the quench amplitude and $\theta$ is the angle between $\Delta\boldsymbol{\lambda}$ and $\boldsymbol{v}_{\parallel}$. Only the heat-engine region satisfying $W_{\mathrm{out}} > 0$, $Q_h > 0$, and $Q_c < 0$ is displayed.

Figure~\ref{a2}(a) shows the contour plot of the exact output work $W_{\mathrm{out}}$. The largest work values are concentrated near the radial direction $\theta = 0$, consistent with $b_{\parallel} \neq 0$ and $b_{\perp} = 0$. At a fixed quench amplitude $\xi$, tilting the quench away from the radial direction increases its transverse component and reduces $W_{\mathrm{out}}$. This trend is consistent with Eq.~\eqref{workt}, in which the transverse component contributes only a quadratic response cost. Figure~\ref{a2}(a) therefore shows that the maximum exact output work within the displayed heat-engine region occurs at $\theta=0$ and a finite quench amplitude.

The corresponding efficiency contour plot is shown in Fig.~\ref{a2}(b). High-efficiency values are also concentrated near the radial direction, indicating that quenches close to this direction are more favorable for efficient energy conversion. The maximum exact output work occurs at an intermediate quench amplitude, whereas larger radial amplitudes can yield higher efficiencies within the positive-work region. This shows a trade-off between $W_{\rm out}$ and $\eta$: the amplitude that maximizes $W_{\mathrm{out}}$ generally differs from that which maximizes $\eta$. For the temperatures considered here, the Carnot efficiency is $\eta_{\mathrm{C}} = 1 - \beta_h/\beta_c = 0.7$.
The exact efficiency remains below this bound throughout the displayed heat-engine region. Values approaching $\eta_{\mathrm{C}}$ occur near the boundary of the operating region rather than at the point of maximum output work. Overall, the two contour plots show that transverse quench components reduce the output work and efficiency, whereas radial quenches provide the more favorable heat-engine operation.

\begin{figure}
\centering
\begin{tabular}{@{}c@{\hspace{0.6in}}c@{}}
\hspace{0.15in}
\begin{overpic}[width=3.3cm]{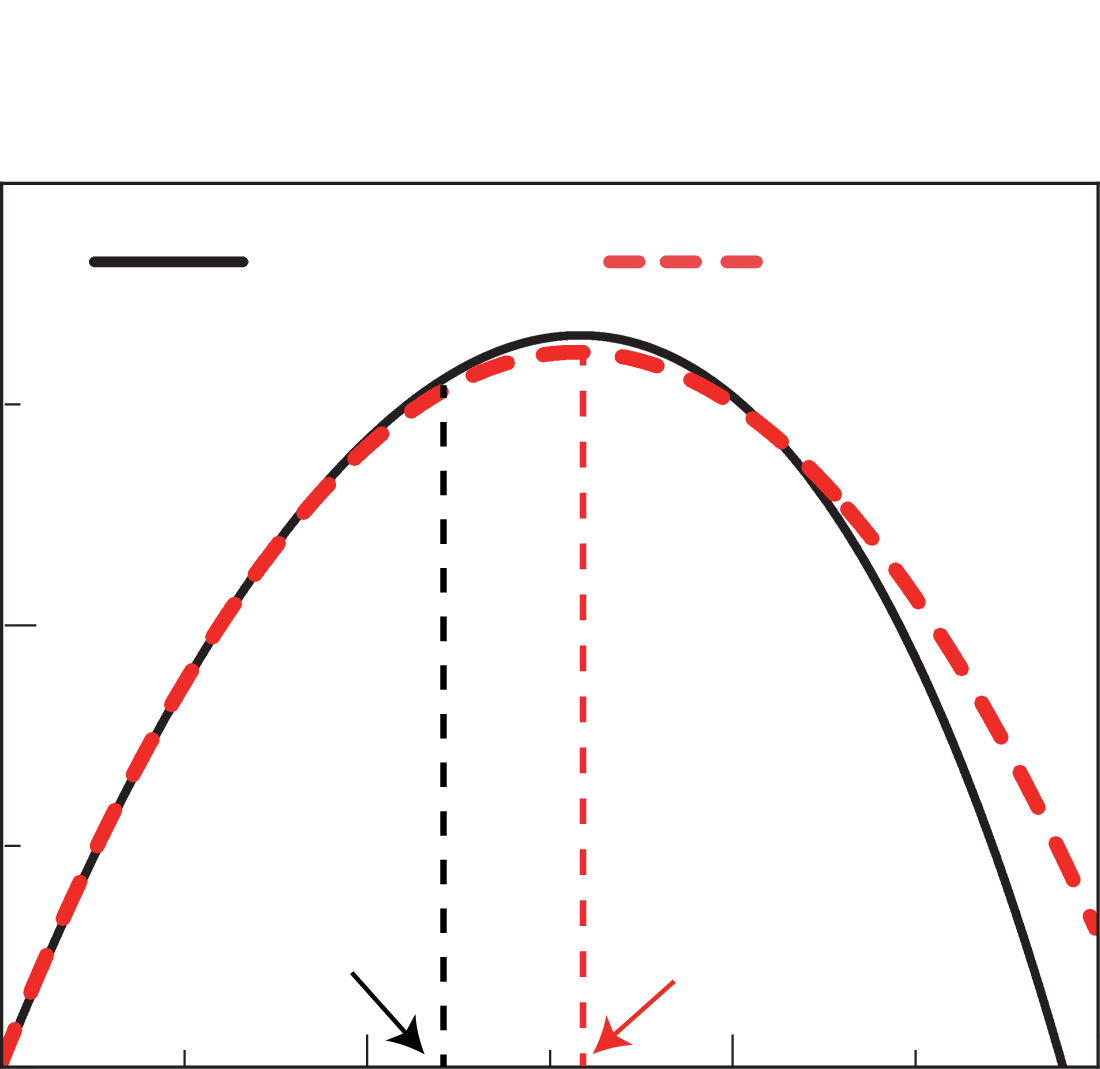}
\put(-22,90){\scalebox{0.9}{(a)}}
\put(-1,-7){\scalebox{0.9}{0}}
\put(27,-7){\scalebox{0.9}{0.5}}
\put(24,8){\scalebox{0.9}{0.6}}
\put(60,8){\scalebox{0.9}{0.79}}
\put(61,-7){\scalebox{0.9}{1.0}}
\put(93,-7){\scalebox{0.9}{1.5}}
\put(-5.5,-2){\scalebox{0.9}{0}}
\put(-12.5,38){\scalebox{0.9}{0.2}}
\put(-12.5,77){\scalebox{0.9}{0.4}}
\put(25,71){\scalebox{0.8}{$W_{\rm out}$}}
\put(73,71){\scalebox{0.8}{$W_{\rm loc}$}}
\put(48,-16){\scalebox{1.1}{$\xi$}}
\end{overpic}
\hspace{0.3in}
\begin{overpic}[width=3.3cm]{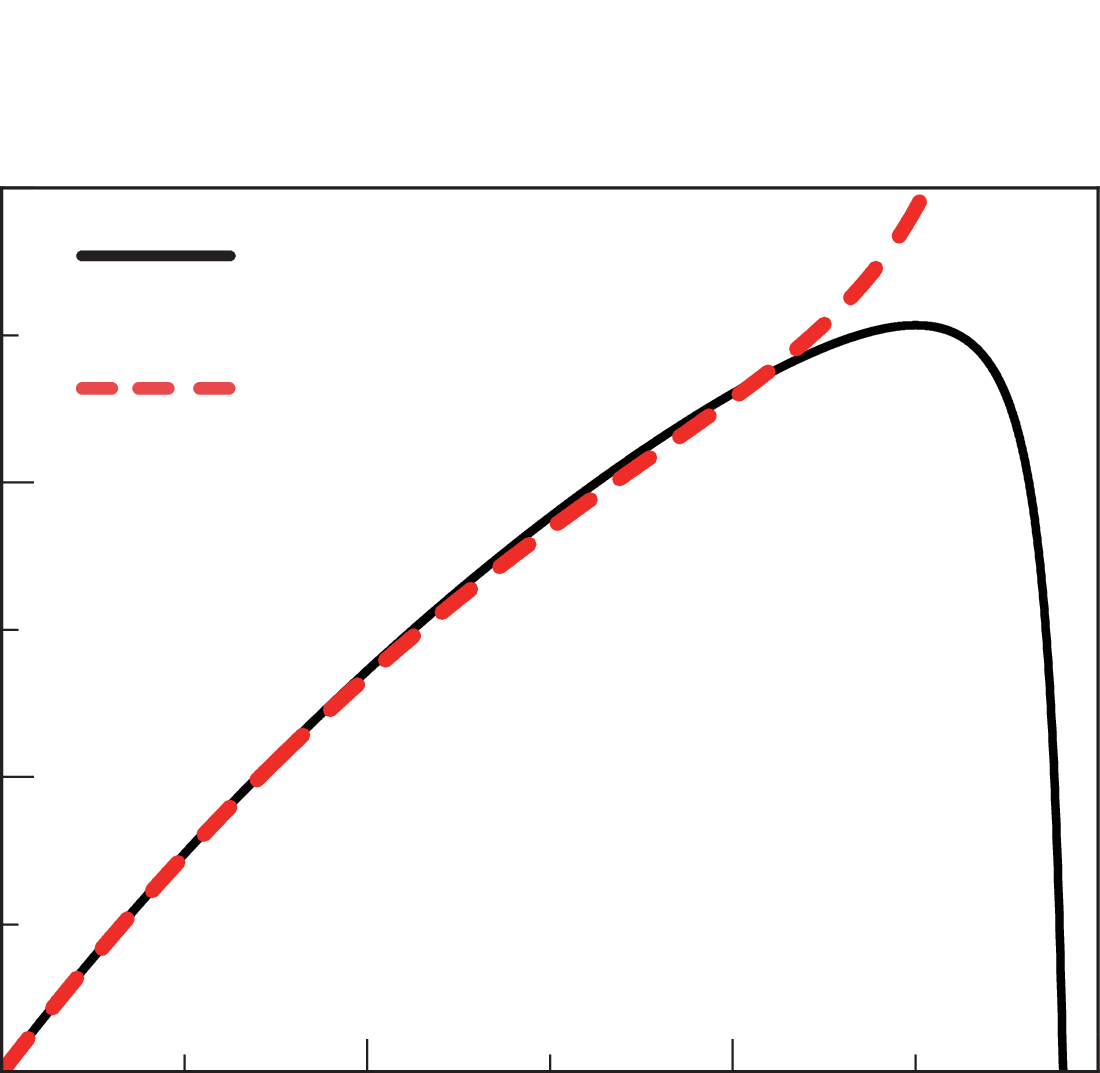}
\put(-22,90){\scalebox{0.9}{(b)}}
\put(-1,-7){\scalebox{0.9}{0}}
\put(27,-7){\scalebox{0.9}{0.5}}
\put(61,-7){\scalebox{0.9}{1.0}}
\put(93,-7){\scalebox{0.9}{1.5}}
\put(-5.5,-2){\scalebox{0.9}{0}}
\put(-12.5,25){\scalebox{0.9}{0.2}}
\put(-12.5,52){\scalebox{0.9}{0.4}}
\put(-12.5,77){\scalebox{0.9}{0.6}}
\put(25,73){\scalebox{0.8}{$\eta$}}
\put(25,60){\scalebox{0.8}{$\eta_{\rm dig}$}}
\put(48,-16){\scalebox{1.1}{$\xi$}}
\end{overpic}
\\[0.2in]
\end{tabular}

\caption{(a) Exact output work $W_\mathrm{out}$ and local quadratic work $W_\mathrm{loc}$, and (b) exact efficiency $\eta$ and diagnostic quantity $\eta_\mathrm{dig}$, along the unconstrained-optimal quench path $\Delta\bm{\lambda}_\mathrm{unc}(\xi) = \xi\bm{v}_\mathrm{unc}$. The black and red vertical dashed lines indicate the selected amplitude bound $\epsilon=0.6$ and the unconstrained optimal amplitude $\xi_{\rm unc}\simeq0.79$, respectively. The other parameters are the same as in Table~\ref{tab}.}
\label{a3}
\end{figure}

\begin{figure*}[t]
\centering
\begin{tabular}{@{}c@{\hspace{0.6in}}c@{}}
\hspace{0.2in}
\begin{overpic}[width=4.5cm]{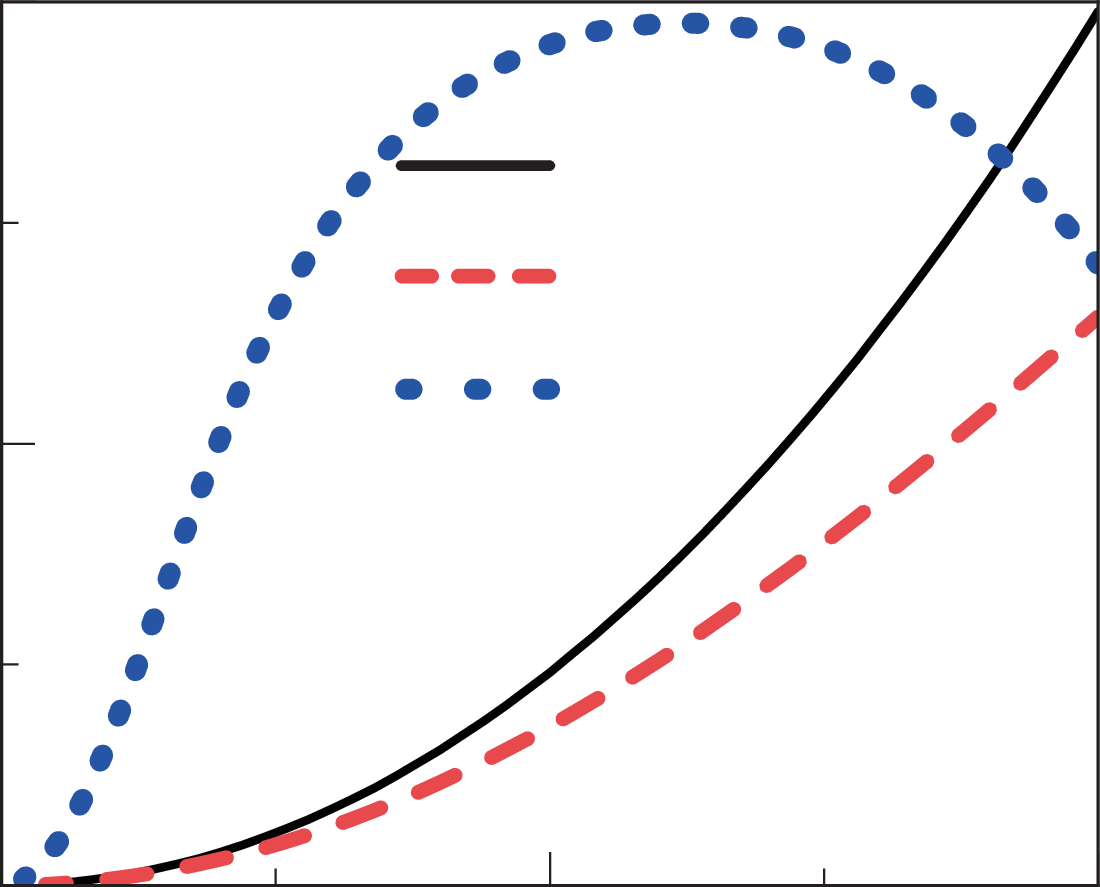}
\put(-22,90){\scalebox{0.9}{(a)}}
\put(-1,-7){\scalebox{0.9}{0}}
\put(45,-7){\scalebox{0.9}{0.25}}
\put(95,-7){\scalebox{0.9}{0.5}}
\put(-5,-2){\scalebox{0.9}{0}}
\put(-13,38.5){\scalebox{0.9}{0.01}}
\put(-13,78){\scalebox{0.9}{0.02}}
\put(52,64){\scalebox{0.9}{$\Delta\bm \lambda_{\rm prop}$}}
\put(52,53.5){\scalebox{0.9}{$\Delta\bm \lambda_{\rm unc}$}}
\put(52,43.5){\scalebox{0.9}{$\Delta\bm \lambda_{\rm opt}$}}
\put(48,-16){\scalebox{1.1}{$\xi$}}
\put(-22,35){\rotatebox{90}{\scalebox{0.9}{$\delta_W$}}}
\end{overpic}
\hspace{0.55in}
\begin{overpic}[width=4.5cm]{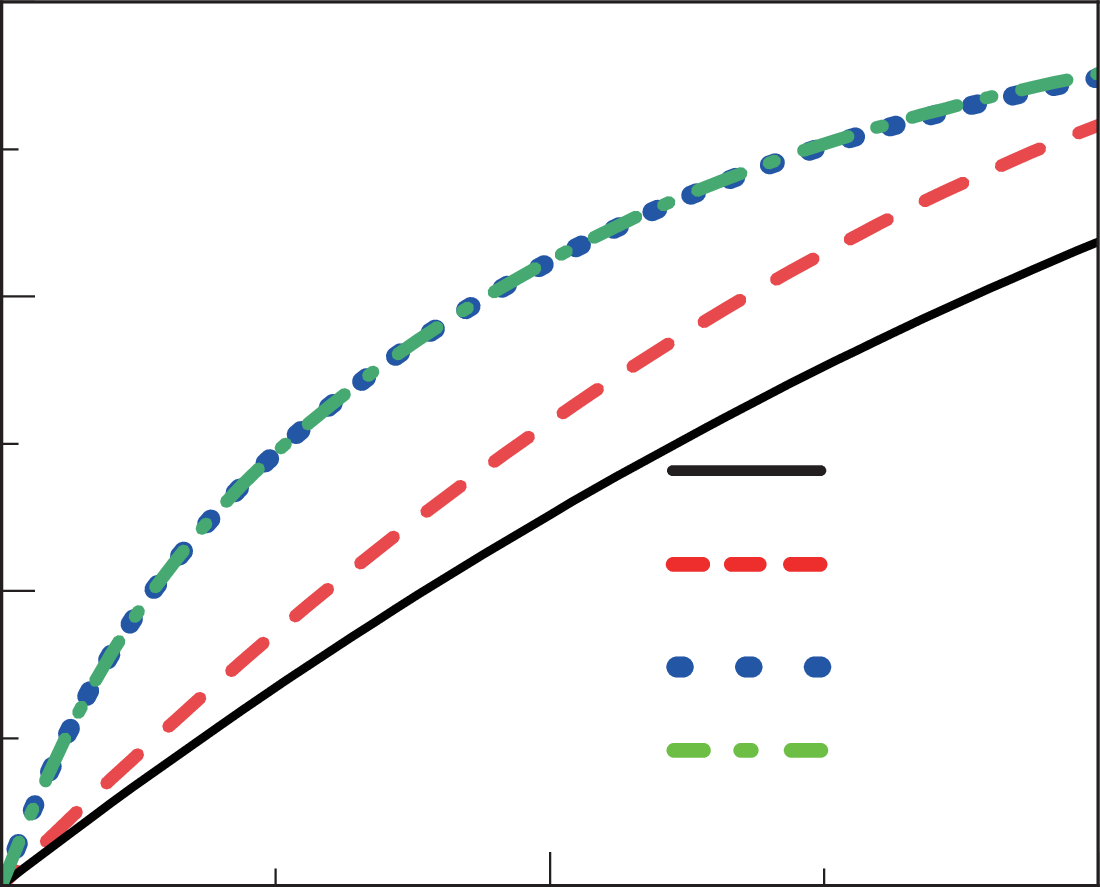}
\put(-22,90){\scalebox{0.9}{(b)}}
\put(-1,-7){\scalebox{0.9}{0}}
\put(45,-7){\scalebox{0.9}{0.25}}
\put(95,-7){\scalebox{0.9}{0.5}}
\put(-5,-2){\scalebox{0.9}{0}}
\put(-13,25){\scalebox{0.9}{0.11}}
\put(-13,51.5){\scalebox{0.9}{0.22}}
\put(-13,78){\scalebox{0.9}{0.33}}
\put(77,36.5){\scalebox{0.9}{$\Delta\bm \lambda_{\rm prop}$}}
\put(77,27.5){\scalebox{0.9}{$\Delta\bm \lambda_{\rm unc}$}}
\put(77,18.5){\scalebox{0.9}{$\Delta\bm \lambda_{\rm opt}$}}
\put(77,10){\scalebox{0.9}{$\Delta\bm \lambda^{\rm ext}_{\rm opt}$}}
\put(48,-16){\scalebox{1.1}{$\xi$}}
\put(-22,33){\rotatebox{90}{\scalebox{0.9}{$W_{\rm out}$}}}
\end{overpic}
\hspace{0.55in}
\begin{overpic}[width=4.5cm]{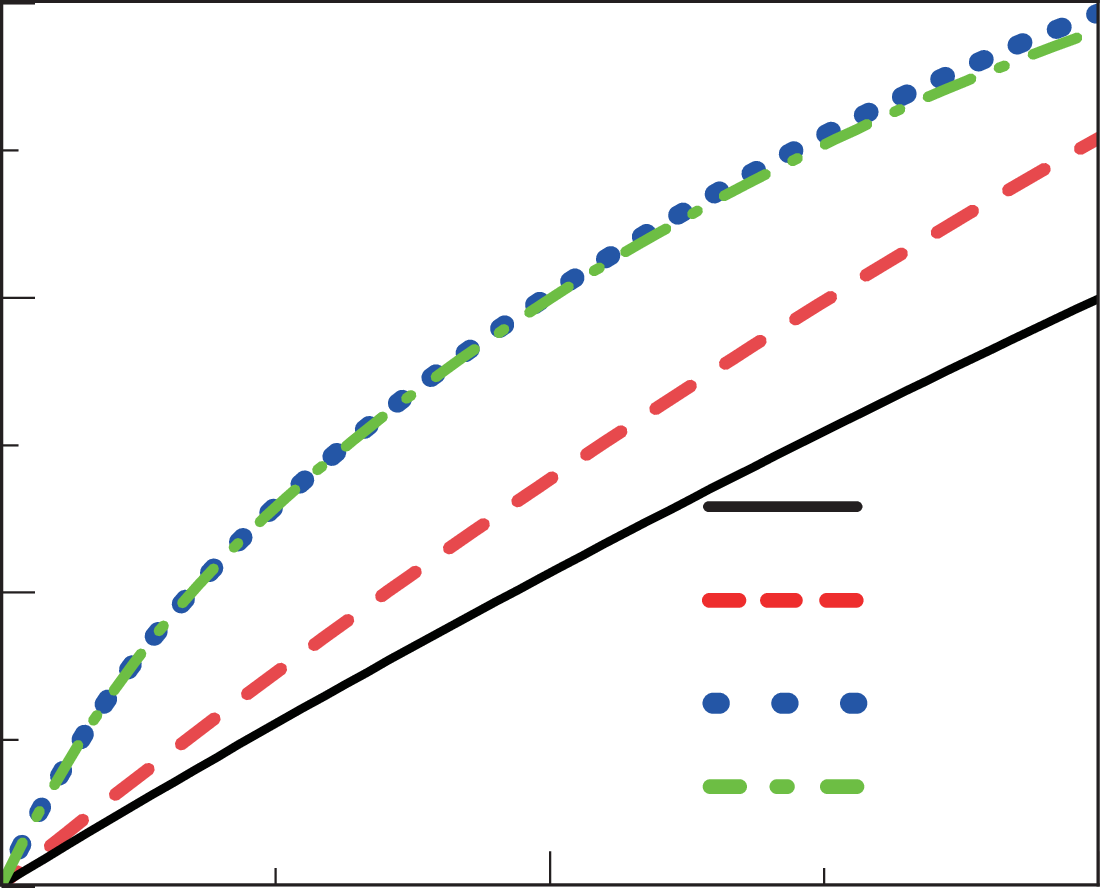}
\put(-23,90){\scalebox{0.9}{(c)}}
\put(-1,-7){\scalebox{0.9}{0}}
\put(45,-7){\scalebox{0.9}{0.25}}
\put(95,-7){\scalebox{0.9}{0.5}}
\put(-5,-2){\scalebox{0.9}{0}}
\put(-13,25){\scalebox{0.9}{0.12}}
\put(-13,51.5){\scalebox{0.9}{0.24}}
\put(-13,78){\scalebox{0.9}{0.36}}
\put(79,33){\scalebox{0.9}{$\Delta\bm \lambda_{\rm prop}$}}
\put(79,24.5){\scalebox{0.9}{$\Delta\bm \lambda_{\rm unc}$}}
\put(79,15){\scalebox{0.9}{$\Delta\bm \lambda_{\rm opt}$}}
\put(79,7.5){\scalebox{0.9}{$\Delta\bm \lambda^{\rm ext}_{\rm opt}$}}
\put(-20,37){\rotatebox{90}{\scalebox{1.1}{$\eta$}}}
\put(48,-16){\scalebox{1.1}{$\xi$}}
\end{overpic}

\\[0.2in]
\end{tabular}
\caption{  (a) Relative deviation \(\delta W\) between \(W_{\rm out}\)
and \(W_{\rm loc}\) along the three  quench paths
\(\Delta\bm\lambda_{\rm prop}(\xi)\),
\(\Delta\bm\lambda_{\rm unc}(\xi)\), and
\(\Delta\bm\lambda_{\rm opt}(\xi)\). (b) Exact output work along these
paths, together with the maximum exact output work. (c) Exact efficiency along the same paths and along
\(\Delta\bm\lambda^{\rm ext}_{\rm opt}(\xi)\). Black solid, red dashed, and blue dotted curves denote
\(\Delta\bm\lambda_{\rm prop}(\xi)\),
\(\Delta\bm\lambda_{\rm unc}(\xi)\), and
\(\Delta\bm\lambda_{\rm opt}(\xi)\), respectively. In panels (b) and
(c), the cyan dash-dotted curve denotes the result associated with
\(\Delta\bm\lambda^{\rm ext}_{\rm opt}(\xi)\). 
The other parameters are the same as in Table~\ref{tab}.
}
\label{a4}
\end{figure*}

\subsection{Interacting XYZ spin chain working medium}
\label{sec32}
We apply the local quadratic work formalism to an interacting spin-$1/2$ XYZ spin chain in a longitudinal field \cite{Hag05,Shi19,Zha22},
\begin{equation}
\hat H_{\rm XYZ}
=
J_x\hat O_x+J_y\hat O_y+J_z\hat O_z+h\hat O_h ,
\end{equation}
where $\hat O_\alpha=\sum_{j=1}^{N}\hat\sigma_j^\alpha\hat\sigma_{j+1}^\alpha \quad (\alpha=x,y,z),$ and 
$\hat O_h=\sum_{j=1}^{N}\hat\sigma_j^z$. Periodic boundary conditions are imposed, with
$\hat{\sigma}_{N+1}^{\alpha}=
\hat{\sigma}_{1}^{\alpha}$. The control vector is $\bm\lambda=(J_x,J_y,J_z,h)^T,$
so the model provides a four-dimensional control space. For a given reference point \(\bm\lambda_0\), the thermal-bias vector
\(\bm b_0\) is obtained from the difference between the hot and cold
thermal expectation values, while the response-cost matrix \(K\) is
computed from Eq.~(\ref{Kcorrelation}). Diagonalizing \(K\) gives four
response eigenchannels.

Table~\ref{tab} lists the local response quantities for each eigenchannel. Mode 2 gives $W_{2,\text{loc}}^{\text{max}}$, which is larger than the corresponding contributions of the other modes. The ordering of $W_{k,\text{loc}}^{\text{max}}$ is not determined by the thermal-bias projection alone. Although mode 4 has the largest thermal-bias projection $b_4$, its much larger response cost eigenvalue $\kappa_4$ leads to $W_{4,\text{loc}}^{\text{max}} < W_{2,\text{loc}}^{\text{max}}$.

To isolate the contribution to $\kappa_k$
 arising from control components that do not commute with the reference Hamiltonian, we decompose $\kappa_k=\kappa_k^{\rm com}+\kappa_k^{\rm nc}$. Here, $\kappa_k^{\rm com}$ and $\kappa_k^{\rm nc}$ are obtained from Eq.~(\ref{cost1}) by restricting the sum to matrix elements connecting states within the same energy eigenspace and states belonging to distinct energy eigenspaces, respectively.

The fraction of $\kappa_k$ arising from control components that do not commute with the reference Hamiltonian,
$q_k^{\rm nc}=\kappa_k^{\rm nc}/\kappa_k$, is also listed in
Table~\ref{tab}.  For mode 3, $q_3^{\mathrm{nc}} \simeq 0.959$, indicating that almost all of $\kappa_3$ arises from the noncommuting contribution. For mode 4, $q_4^{\mathrm{nc}} \simeq 0.064$, so most of  $\kappa_4$  arises from the commuting contribution. These quantities are obtained by considering each eigenchannel separately. By contrast, the quench vector that maximizes $W_{\rm loc}$ generally combines components from several eigenchannels.

For the XYZ spin chain, the resulting quench vector is
\begin{equation}
\Delta\bm{\lambda}_{\rm unc}
=\frac{1}{2}\sum_{k=1}^{4}
\frac{b_k}{\kappa_k}\bm{v}_k .
\label{dlam}
\end{equation}

We denote the amplitude of $\Delta\boldsymbol{\lambda}_{\rm unc}$ by
$\xi_{\rm unc} = \Vert\Delta\boldsymbol{\lambda}_{\rm unc}\Vert$
and its unit direction by
$\boldsymbol{v}_{\rm unc} = \Delta\boldsymbol{\lambda}_{\rm unc}/\xi_{\rm unc}.$
Varying the amplitude along $\boldsymbol{v}_{\rm unc}$  defines the unconstrained quench path
$\Delta\boldsymbol{\lambda}_{\rm unc}(\xi) = \xi \boldsymbol{v}_{\rm unc}.$
For the parameters listed in Table~\ref{tab}, $\xi_{\rm unc}\simeq0.79$ and
$\bm{v}_{\rm unc}\simeq(0.133,0.537,0.081,0.829)$, with the components ordered as $(J_x,J_y,J_z,h)$.

Figure~\ref{a3}(a) compares the exact output work $W_{\rm out}$ with the local quadratic work $W_{\rm loc}$ along the unconstrained quench path $\Delta\boldsymbol{\lambda}_{\rm unc}(\xi)$. At small amplitudes, $W_{\rm loc}$ closely agrees with $W_{\rm out}$. As the amplitude increases, the difference between them grows because the higher-order terms omitted from $W_{\rm loc}$ become more important. Using $\delta W \leq 0.02$ as a practical criterion for quantitative agreement, we find that the relative deviation remains below $2\%$ for $\xi \lesssim 0.6$. Since $\xi_{\rm unc} \simeq 0.79$ exceeds this upper limit, we choose a smaller amplitude bound $\epsilon \leq 0.6$ and maximize $W_{\rm loc}$ over quench vectors satisfying $\Vert\Delta\boldsymbol{\lambda}\Vert \leq \epsilon$.

In Fig. \ref{a3}(b), we show the corresponding comparison between the exact efficiency $\eta$ and the diagnostic quantity $\eta_{\rm dig}$. Their close agreement over the same small-amplitude interval along
this path follows from the accuracy of the quadratic work
approximation, since both quantities use the same exact energy input
$Q_h$.

We therefore set the maximum quench amplitude to $\epsilon = 0.6$ and let $\xi$ vary within $0 < \xi \leq \epsilon$. For each value of $\xi$, we maximize $W_{\rm loc}$ over all quench vectors satisfying $\Vert\Delta\boldsymbol{\lambda}\Vert \leq \xi$. Using Eq.~\eqref{con1}, the resulting quench vector is
\begin{equation}
\Delta\bm{\lambda}_{\rm opt}(\xi)
=\frac{1}{2}
\sum_{k=1}^4
\frac{b_k}{\kappa_k+\zeta(\xi)}
\bm v_k,
\label{optlam}
\end{equation}
where $\zeta(\xi) > 0$ is determined as the unique solution of $\sum_{k=1}^4
b_k^2/[4(\kappa_k+\zeta(\xi))^2]
=\xi^2.$
This condition ensures that
$\Vert\Delta\boldsymbol{\lambda}_{\rm opt}(\xi)\Vert = \xi.$
As $\xi$ varies over $0 < \xi \leq \epsilon$, these quench vectors define the constrained quench path $\Delta\boldsymbol{\lambda}_{\rm opt}(\xi)$. The unit direction along this path is
$\boldsymbol{v}_{\rm opt}(\xi) = \Delta\boldsymbol{\lambda}_{\rm opt}(\xi)/\xi.$
Because $\zeta(\xi)$ changes with $\xi$, the relative weights of the four response eigenchannels also change. Therefore, the direction $\boldsymbol{v}_{\rm opt}(\xi)$ generally varies with the quench amplitude.

For comparison, we also introduce the proportional quench path
$\Delta\bm{\lambda}_{\rm prop}(\xi)=\xi\bm v_{\rm prop}$, where
$\bm v_{\rm prop}=\bm\lambda_0/|\bm\lambda_0|$.
For the present reference point $\bm \lambda_0$, the unit direction is 
$\bm v_{\rm prop}\simeq(0.054,0.436,0.218,0.872)$, with the components
ordered as $(J_x,J_y,J_z,h)$. This proportional quench path preserves the relative
proportions of the control parameters and does not use the response
information contained in  $K$ and $\bm b_0$. 

Furthermore, to identify a common amplitude interval for the following pathwise
comparisons, we evaluate $\delta W$ along the proportional, unconstrained, and constrained quench paths.
Figure~\ref{a4}(a) shows that the relative deviations along all three paths do not exceed $0.02$ for $0 < \xi \leq 0.5$. We therefore use $\xi = 0.5$ as the common upper bound for the comparisons between paths. To benchmark these three quench paths, we further determine, for each $0<\xi\leq0.5$, the optimal quench vector $\Delta\bm\lambda^{\rm ext}_{\rm opt}(\xi)$ by maximizing exact output work $W_{\rm out}$ over all quench directions at the same amplitude (see Appendix \ref{appE} for details).

Figure~\ref{a4}(b) compares four curves for the exact output work within the common interval $0<\xi\leq0.5$. The black, red, and blue curves correspond to the three prescribed quench paths
\(\Delta\boldsymbol{\lambda}_{\rm prop}(\xi)\),
\(\Delta\boldsymbol{\lambda}_{\rm unc}(\xi)\), and
\(\Delta\boldsymbol{\lambda}_{\rm opt}(\xi)\), respectively.
Among these three paths, the constrained quench path yields the largest exact output work for all plotted values of \(\xi\), followed by the unconstrained and proportional quench paths. The cyan curve shows
the maximum exact output work obtained by searching over all quench
directions at each amplitude \(\xi\). It nearly coincides  with the constrained quench path over $0<\xi\le0.5$. This numerical agreement shows that evaluating $W_{\rm out}$ at the quench vector that maximizes $W_{\rm loc}$ gives a value close to the maximum exact output work throughout this interval.

Figure~\ref{a4}(c) shows the corresponding exact efficiencies. Among the three quench paths for \(\Delta\boldsymbol{\lambda}_{\rm prop}(\xi)\),
\(\Delta\boldsymbol{\lambda}_{\rm unc}(\xi)\), and
\(\Delta\boldsymbol{\lambda}_{\rm opt}(\xi)\), the constrained quench path
gives the highest efficiency throughout the common comparison
interval $0<\xi\leq0.5$. We also plot the efficiency evaluated along the quench vector that maximizes the exact output work.  This curve almost coincides with the constrained quench path efficiency
at small \(\xi\), while at larger \(\xi\) the constrained quench path
gives a slightly higher efficiency. Overall, the local response theory provides an effective way to
select multiparameter quench vectors that nearly attain the maximum
exact output work over the small-amplitude interval examined here. The selected
quench vector also exhibits a comparable exact efficiency, confirming its favorable heat-engine performance in the full exact cycle.

\section{DISCUSSION AND CONCLUSIONS}
\label{con}
In this work, we developed a local response theory for selecting small multiparameter quenches in sudden quench quantum Otto cycles. Expanding the exact output work around a common reference point yields a local quadratic work composed of a linear thermal-bias gain and a quadratic response cost. In the energy eigenbasis of the reference Hamiltonian, the gain depends only on the diagonal matrix elements of the quench operator, whereas the cost contains both diagonal and off diagonal contributions. Diagonalizing the response cost matrix resolves the local quadratic work into response eigenchannels characterized by the thermal-bias projection $b_k$  and the response cost eigenvalue $\kappa_k$. This representation makes explicit how the balance between work gain and response cost determines the optimal combination of available controls.

Once the control metric $G$ and the amplitude bound $\epsilon$ are specified, the theory determines the quench vector that maximizes local quadratic work
 over the admissible region $B_{\epsilon}$. Although this vector is obtained from a local approximation, its exact output work falls short of the maximum exact output work over the same region by at most a term of order $\epsilon^3$. The local response theory therefore provides a constructive route from equilibrium information at a single reference point to near optimal exact output work in the small amplitude limit.

The present work therefore provides a constructive criterion for selecting a small multiparameter quench by maximizing the local quadratic work. This complements Ref.~\cite{Wat26}, which compares the performance of simultaneous multiparameter quenches with that of the corresponding single-parameter cycles for fixed endpoint parameters.  For the two-level working medium, the response eigenchannels reduce to radial and transverse directions. The radial channel carries the thermal-bias gain, whereas the transverse channel contributes only a quadratic response cost. In the interacting finite XYZ spin chain, the optimized quench vector combines several response eigenchannels and yields near-maximum exact output work when compared with the full directional search over the amplitude interval examined numerically. Its exact efficiency also remains favorable, although efficiency is not optimized.

The present analysis assumes linearly controlled finite dimensional working media, complete thermalization during the isochoric strokes, instantaneous work strokes, and small quenches. Its quantitative accuracy is controlled by the neglected third and higher order terms and must therefore be assessed over the amplitude range of interest. In addition, the control metric $G$ and the amplitude bound $\epsilon$ must be specified as part of the physical control protocol. Possible extensions include local expansions of heat and efficiency, incomplete thermalization during finite time isochoric strokes, nonlinear control dependence, and noninstantaneous work strokes.

\begin{acknowledgements}
This work was supported by the National Natural Science Foundation of China under Grant No. 12465009, the Major Program of the Jiangxi Provincial Natural Science Foundation under Grant No. 20224ACB201007, and the Natural Science Foundation Innovation-Development Joint Program of Hubei Province under Grant No. 2026AFC0003.
\end{acknowledgements}
\appendix

\section{Bogoliubov--Kubo--Mori covariance form of the response matrix}
\label{appA}
In this appendix, we derive the relation $D^{(\beta)} = -\beta C^{(\beta)}$ between the static response matrix and the BKM covariance matrix, and prove that $C^{(\beta)}$ is positive semidefinite. 
We consider the Gibbs state $\hat{\rho}_\beta(\boldsymbol{\lambda}) = e^{-\beta \hat{H}(\boldsymbol{\lambda})} / Z_\beta(\boldsymbol{\lambda})$, with $Z_\beta(\boldsymbol{\lambda}) = \operatorname{Tr} [e^{-\beta \hat{H}(\boldsymbol{\lambda})}]$. The Hamiltonian is assumed to depend linearly on the control parameters, $\hat{H}(\boldsymbol{\lambda}) = \hat{H}_0 + \sum_\mu \lambda_\mu \hat{O}_\mu$, with $\hat{O}_\mu$ independent of $\boldsymbol{\lambda}$.

The thermal expectation values of the controlled operators are collected in the vector
$\langle \hat{\boldsymbol{O}} \rangle_{\beta,\boldsymbol{\lambda}}$, whose components are $\langle \hat{O}_\mu \rangle_{\beta,\boldsymbol{\lambda}} = \operatorname{Tr}\big[\hat{\rho}_\beta(\boldsymbol{\lambda}) \hat{O}_\mu\big].$ The static response matrix $D^{(\beta)}$ is defined by its matrix elements
\begin{equation}
\left(D^{(\beta)}\right)_{\mu\nu}
=
\frac{\partial
\langle \hat O_{\mu}\rangle_{\beta,\bm{\lambda}}}
{\partial \lambda_{\nu}}
=
{\rm Tr}\left[
\frac{\partial \hat{\rho}_{\beta} (\bm{\lambda})}
{\partial \lambda_{\nu}}
\hat O_{\mu}
\right],
\label{duv}
\end{equation}
where the last equality follows from \(\partial_{\lambda_\nu}\hat O_\mu=0\).

To differentiate the operator exponential, we use the standard integral formula for parameter differentiation of exponential operators \cite{Hub57,Wil67}
\begin{equation}
\frac{\partial e^{-\beta\hat H}}{\partial\lambda_\nu}
=
-\beta
\int_0^1 ds\,
e^{-\beta(1-s)\hat H}
\hat O_\nu
e^{-\beta s\hat H},
\end{equation}
which is valid even when $\hat{H}$ and $\hat{O}_\nu$ do not commute.
For the normalized Gibbs state
\(\hat\rho_\beta\), differentiation gives
\begin{align}
\frac{\partial \hat\rho_\beta}{\partial\lambda_\nu}
&=
\frac{1}{Z_\beta}
\frac{\partial e^{-\beta\hat H}}{\partial\lambda_\nu}
-
\frac{e^{-\beta\hat H}}{Z_\beta^2}
\frac{\partial Z_\beta}{\partial\lambda_\nu}
\nonumber\\
&=
-\beta
\int_0^1 ds\,
\frac{
e^{-\beta(1-s)\hat H}
\hat O_\nu
e^{-\beta s\hat H}
}
{Z_\beta}
-
\hat\rho_\beta
\frac{1}{Z_\beta}
\frac{\partial Z_\beta}{\partial\lambda_\nu}
\nonumber\\
&=-\beta
\int_0^1 ds\,
\frac{
e^{-\beta(1-s)\hat H}
\hat O_\nu
e^{-\beta s\hat H}
}
{Z_\beta^{1-s}Z_\beta^{s}}
-
\hat\rho_\beta
\frac{1}{Z_\beta}
\frac{\partial Z_\beta}{\partial\lambda_\nu}
\nonumber\\
&=-\beta\int_0^1 ds\,\hat\rho_\beta^{1-s}
\hat O_\nu
\hat\rho_\beta^s-
\hat\rho_\beta
\frac{1}{Z_\beta}
\frac{\partial Z_\beta}{\partial\lambda_\nu}.
\label{rhod}
\end{align}

The derivative of the partition function is
\begin{align}
\frac{\partial Z_\beta}{\partial\lambda_\nu}
&=
{\rm Tr}
\left[
\frac{\partial e^{-\beta\hat H}}{\partial\lambda_\nu}
\right]
\nonumber\\
&=
-\beta
\int_0^1 ds\,
{\rm Tr}
\left[
e^{-\beta(1-s)\hat H}
\hat O_\nu
e^{-\beta s\hat H}
\right]
\nonumber\\
\nonumber&=
-\beta
{\rm Tr}
\left[
e^{-\beta\hat H}\hat O_\nu
\right]
\nonumber\\
&=-\beta Z_\beta
\langle \hat O_\nu\rangle_{\beta,\bm \lambda}.
\label{partition}
\end{align}

In the third line, the cyclic invariance of the trace removes the s-dependence of the integrand.

Substituting Eq.~\eqref{partition} into Eq.~\eqref{rhod}, we obtain
\begin{align}
\frac{\partial \hat\rho_\beta}{\partial\lambda_\nu}
&=
-\beta
\int_0^1 ds\,
\hat\rho_\beta^{1-s}
\hat O_\nu
\hat\rho_\beta^s
+
\beta
\hat\rho_\beta
\langle \hat O_\nu\rangle_{\beta,\bm \lambda}
\nonumber\\
&=
-\beta
\int_0^1 ds\,
\hat\rho_\beta^{1-s}
\hat O_\nu
\hat\rho_\beta^s
+
\beta
\int_0^1 ds\,
\hat\rho_\beta^{1-s}
\langle \hat O_\nu\rangle_{\beta,\bm \lambda}
\hat\rho_\beta^s
\nonumber\\
&=
-\beta
\int_0^1 ds\,
\hat\rho_\beta^{1-s}
\left(
\hat O_\nu-\langle \hat O_\nu\rangle_{\beta,\bm \lambda}
\right)
\hat\rho_\beta^s
\nonumber\\
&=
-\beta
\int_0^1 ds\,
\hat\rho_\beta^{1-s}
\delta\hat O_\nu
\hat\rho_\beta^s,
\label{rhod1}
\end{align}
where \(\delta\hat{O}_\nu = \hat{O}_\nu - \langle\hat{O}_\nu\rangle_{\beta,\bm\lambda}\hat{I}\)  denotes the fluctuation of \(\hat{O}_\nu\) about its thermal expectation value.

Substituting Eq.~\eqref{rhod1} into the definition of \(\left(D^{(\beta)}\right)_{\mu\nu}\), Eq.~\eqref{duv} can be written as 
\begin{align}
\left(D^{\beta}\right)_{\mu\nu}
=&
-\beta
\int_0^1 ds\,
{\rm Tr}
\left[
\hat\rho_\beta^{1-s}
\delta\hat O_\nu
\hat\rho_\beta^{s}
\hat O_\mu
\right]
\nonumber\\
=&
-\beta
\int_0^1 ds\,
{\rm Tr}
\left[
\hat\rho_\beta^{1-s}
\delta\hat O_\nu
\hat\rho_\beta^{s}
\left(
\delta\hat O_\mu
+
\langle \hat O_\mu\rangle_{\beta,\bm\lambda}
\hat{I}\right)
\right]
\nonumber\\
=&
-\beta
\int_0^1 ds\,
{\rm Tr}
\left[
\hat\rho_\beta^{1-s}
\delta\hat O_\nu
\hat\rho_\beta^{s}
\delta\hat O_\mu
\right]
\nonumber\\
&-\beta
\langle \hat O_\mu\rangle_{\beta,\bm\lambda}
\int_0^1 ds\,
{\rm Tr}
\left[
\hat\rho_\beta^{1-s}
\delta\hat O_\nu
\hat\rho_\beta^{s}
\right]
\nonumber\\
=&
-\beta
\int_0^1 ds\,
{\rm Tr}
\left[
\hat\rho_\beta^{1-s}
\delta\hat O_\nu
\hat\rho_\beta^{s}
\delta\hat O_\mu
\right]
\nonumber\\
&-\beta
\langle \hat O_\mu\rangle_{\beta,\bm\lambda}
\int_0^1 ds\,
{\rm Tr}
\left[
\hat\rho_\beta
\delta\hat O_\nu
\right]
\nonumber\\
=&
-\beta
\int_0^1 ds\,
{\rm Tr}
\left[
\hat\rho_\beta^{1-s}
\delta\hat O_\nu
\hat\rho_\beta^{s}
\delta\hat O_\mu
\right]
\nonumber\\
=&
-\beta
\int_0^1 ds\,
{\rm Tr}
\left[
\hat\rho_\beta^{s}
\delta\hat O_\mu
\hat\rho_\beta^{1-s}
\delta\hat O_\nu
\right].
\label{KMcov}
\end{align}

In the fourth line, the cyclic invariance of the trace has been used to write ${\rm Tr}\left[ \hat\rho_\beta^{1-s} \delta\hat O_\nu \hat\rho_\beta^s \right] = {\rm Tr}\left[ \hat\rho_\beta \delta\hat O_\nu \right].$ This term vanishes because \({\rm Tr}[\hat\rho_\beta\delta\hat O_\nu] = \langle\hat O_\nu\rangle_{\beta,\bm\lambda} - \langle\hat O_\nu\rangle_{\beta,\bm\lambda} =0\). The last line again follows from cyclic invariance of the trace.

We define the BKM covariance matrix $C^{(\beta)}$ by its matrix elements $\left(C^{(\beta)}\right)_{\mu\nu}
=\int_0^1 ds\,
\operatorname{Tr}[
\hat{\rho}_{\beta}^{s}
\delta\hat{O}_{\mu}
\hat{\rho}_{\beta}^{1-s}
\delta\hat{O}_{\nu}]$.
Equation~(\ref{KMcov}) then gives
$\left(D^{(\beta)}\right)_{\mu\nu}
=-\beta\left(C^{(\beta)}\right)_{\mu\nu}$.
Equivalently, in matrix form,
\begin{equation}
D^{(\beta)}=-\beta C^{(\beta)}.
\end{equation}

For Hermitian controlled operators, $C^{(\beta)}$ is real
and symmetric. Indeed, changing the integration variable from
$s$ to $1-s$ and using the cyclic invariance of the trace gives
\begin{align}
\left(C^{(\beta)}\right)_{\nu\mu}
&=
\int_0^1 ds\,
\operatorname{Tr}
\left[
\hat{\rho}_{\beta}^{s}
\delta\hat{O}_{\nu}
\hat{\rho}_{\beta}^{1-s}
\delta\hat{O}_{\mu}
\right]
\nonumber\\
&=
\int_0^1 ds\,
\operatorname{Tr}
\left[
\hat{\rho}_{\beta}^{s}
\delta\hat{O}_{\mu}
\hat{\rho}_{\beta}^{1-s}
\delta\hat{O}_{\nu}
\right]
\nonumber\\
&=
\left(C^{(\beta)}\right)_{\mu\nu}.
\end{align}

Moreover, Hermiticity gives
$[\left(C^{(\beta)}\right)_{\mu\nu}]^*=\left(C^{(\beta)}\right)_{\nu\mu}$.
Together with the symmetry relation above, this shows that the matrix elements of $C^{(\beta)}$ are real.


Finally, we show that the BKM covariance matrix $C^{(\beta)}$ is positive semidefinite, which in turn implies the positive semidefiniteness of the response-cost matrix $K$. 
For any real vector $\boldsymbol{x}$, define $\delta\hat{X}=\sum_{\mu}x_{\mu}\delta\hat{O}_{\mu}$. Then
\begin{align}
\bm x^T C^{(\beta)}\bm x
&=
\int_0^1 ds\,
{\rm Tr}\left[
\hat\rho_\beta^s
\delta\hat X
\hat\rho_\beta^{1-s}
\delta\hat X
\right]
\nonumber\\
&=
\int_0^1 ds\,
{\rm Tr}\left[
\hat A_s\hat A_s^\dagger
\right]
\ge 0,
\end{align}
where $\hat A_s=\hat\rho_\beta^{s/2}
\delta\hat X
\hat\rho_\beta^{(1-s)/2}.$ For a full-rank Gibbs state, equality holds only if $\delta \hat{X} = 0$, because $\hat{\rho}_\beta^{s/2}$ and $\hat{\rho}_\beta^{(1-s)/2}$ are invertible. Thus, a vanishing BKM covariance implies that $\hat{X}$ is proportional to the identity. Therefore, \(C^{(\beta)}\) is positive semidefinite. Since $K=[\beta_h C^{(\beta_h)}+\beta_c C^{(\beta_c)}]/2,$ the matrix \(K\) is also positive semidefinite for \(\beta_h,\beta_c>0\).

\section{Spectral representation of the  thermal-bias gain and response cost}
\label{appD}
In this appendix, we express the thermal-bias gain and quadratic response cost in the energy basis of the reference Hamiltonian,
$\hat H_r\equiv\hat H(\bm{\lambda}_0)
=\sum_n E_n|n\rangle\langle n|.$
For degenerate spectra, this decomposition is understood in terms of the spectral projectors, $\hat{H}_r = \sum_E E\Pi_E$. Matrix elements within the same energy eigenspace give the commuting contribution, whereas those connecting distinct energy eigenspaces give the noncommuting contribution.

At inverse temperature \(\beta_\alpha\), with \(\alpha=h,c\), the Gibbs
state at the reference point is
\begin{equation}
\hat\rho_\alpha
=
\frac{e^{-\beta_\alpha \hat H_{\rm r}}}{Z_\alpha}
=
\sum_n p_n^{(\alpha)}
|n\rangle\langle n|,
\end{equation}
where $p_n^{(\alpha)}
=e^{-\beta_\alpha E_n}/Z_\alpha.$
For the local cycle considered in the main text, the two Hamiltonians are
connected by the quench vector
\(\Delta\boldsymbol{\lambda}=\boldsymbol{\lambda}_h-\boldsymbol{\lambda}_c\). Because the Hamiltonian depends linearly on the control parameters, the
corresponding Hamiltonian change is
$\hat H(\boldsymbol{\lambda}_h)-\hat H(\boldsymbol{\lambda}_c)
=
\sum_\mu
(\lambda_{h,\mu}-\lambda_{c,\mu})\hat O_\mu.$
It is therefore convenient to collect this control-space displacement into
a single quench operator,
\begin{equation}
\hat X
=\sum_\mu
\Delta\lambda_\mu \hat O_\mu .
\end{equation}

For each thermal branch $\alpha=h,c$, we define the centered
controlled operators as
$\delta\hat O_{\mu,\alpha}
=\hat O_\mu
-\langle\hat O_\mu\rangle_{\beta_\alpha,\boldsymbol{\lambda}_0}
\hat I.$
The centered quench operator is then
$\delta\hat X_\alpha
=\hat X-\langle\hat X\rangle_{\beta_\alpha,\boldsymbol{\lambda}_0}\hat I
=\sum_\mu
\Delta\lambda_\mu\delta\hat O_{\mu,\alpha}.$

We first write the thermal-bias gain in the same eigenbasis. From the
definition of \(\boldsymbol b_0\), one has
\begin{align}
\Delta\boldsymbol{\lambda}\cdot\boldsymbol b_0
&=
\sum_\mu
\Delta\lambda_\mu
\left(
\langle\hat O_\mu\rangle_{\beta_h}
-
\langle\hat O_\mu\rangle_{\beta_c}
\right)
\nonumber\\
&=
\langle\hat X\rangle_{\beta_h}
-
\langle\hat X\rangle_{\beta_c}
\nonumber\\
&=
{\rm Tr}
\left[
(\hat\rho_h-\hat\rho_c)\hat X
\right]
\nonumber\\
&=
{\rm Tr}
\left[
\sum_n
\left(
p_n^{(h)}-p_n^{(c)}
\right)
|n\rangle\langle n|\hat X
\right]
\nonumber\\
&=
\sum_n
\left(
p_n^{(h)}-p_n^{(c)}
\right)
\left(X\right)_{nn}.
\end{align}

Thus, the thermal-bias gain depends only on the energy-diagonal matrix elements of $\hat{X}$.

Using the bilinearity of the BKM covariance, the quadratic form associated with \(C^{(\beta_\alpha)}\) can be written as
\begin{align}
&
\Delta\boldsymbol{\lambda}^{T}
C^{(\beta_\alpha)}
\Delta\boldsymbol{\lambda}
\nonumber\\
&\quad=
\sum_{\mu,\nu}
\Delta\lambda_\mu
\left(C^{(\beta_\alpha)}\right)_{\mu\nu}
\Delta\lambda_\nu
\nonumber\\
&\quad=
\sum_{\mu,\nu}
\Delta\lambda_\mu\Delta\lambda_\nu
\int_0^1 ds\,
{\rm Tr}
\left[
\hat\rho_\alpha^s
\delta\hat O_{\mu,\alpha}
\hat\rho_\alpha^{1-s}
\delta\hat O_{\nu,\alpha}
\right]
\nonumber\\
&\quad=
\int_0^1 ds\,
{\rm Tr}
\left[
\hat\rho_\alpha^s
(\sum_\mu
\Delta\lambda_\mu
\delta\hat O_{\mu,\alpha})
\hat\rho_\alpha^{1-s}
(\sum_\nu\Delta\lambda_\nu
\delta\hat O_{\nu,\alpha})
\right]
\nonumber\\
&\quad=
\int_0^1 ds\,
{\rm Tr}
\left[
\hat\rho_\alpha^s
\delta\hat X_\alpha
\hat\rho_\alpha^{1-s}
\delta\hat X_\alpha
\right].
\label{dldl}
\end{align}

Since \(\hat\rho_\alpha\) is diagonal in the eigenbasis of
\(\hat H_{\rm r}\), one has
\begin{equation}
\hat\rho_\alpha^s
=
\sum_m
\left(p_m^{(\alpha)}\right)^s
|m\rangle\langle m| .
\label{rhos}
\end{equation}

Substituting Eq. (\ref{rhos}) into Eq. (\ref{dldl}) gives
\begin{align}
&
{\rm Tr}
\left[
\hat\rho_\alpha^s
\delta\hat X_\alpha
\hat\rho_\alpha^{1-s}
\delta\hat X_\alpha
\right]
\nonumber\\
&\quad =
\sum_{m,n}
\left(p_m^{(\alpha)}\right)^s
\left(p_n^{(\alpha)}\right)^{1-s}
\langle m|\delta\hat X_\alpha|n\rangle
\langle n|\delta\hat X_\alpha|m\rangle 
\nonumber\\
&\quad =
\sum_{m,n}
\left(p_m^{(\alpha)}\right)^s
\left(p_n^{(\alpha)}\right)^{1-s}
\left|
(\delta X_\alpha)_{mn}
\right|^2,
\end{align}
where the last equality follows from the Hermiticity of
\(\delta\hat X_\alpha\), namely
\(\langle n|\delta\hat X_\alpha|m\rangle
=\langle m|\delta\hat X_\alpha|n\rangle^{*}\).
 Here, we define the logarithmic mean
\begin{equation}
L(x,y)
=
\int_0^1 ds\,x^s y^{1-s}
=
\begin{cases}
\dfrac{x-y}{\ln x-\ln y}, & x\neq y,\\[6pt]
x, & x=y .
\end{cases}
\end{equation}

Performing the \(s\) integration in Eq. (\ref{dldl}), we obtain
\begin{equation}
\Delta\boldsymbol{\lambda}^{T}
C^{(\beta_\alpha)}
\Delta\boldsymbol{\lambda}
=
\sum_{m,n}
L\!\left(
p_m^{(\alpha)},p_n^{(\alpha)}
\right)
\left|
(\delta X_\alpha)_{mn}
\right|^2 .
\end{equation}

Using
$K=\frac12
\left[\beta_h C^{(\beta_h)}
+\beta_c C^{(\beta_c)}
\right],$ the quadratic response cost becomes
\begin{equation}
\Delta\boldsymbol{\lambda}^{T}
K
\Delta\boldsymbol{\lambda}
=
\frac12
\sum_{\alpha=h,c}
\beta_\alpha
\sum_{m,n}
L\!\left(
p_m^{(\alpha)},p_n^{(\alpha)}
\right)
\left|
(\delta X_\alpha)_{mn}
\right|^2 .
\label{cost}
\end{equation}

Since \(L(x,y)>0\) for \(x,y>0\), Eq. (\ref{cost}) gives a spectral
representation of the positive quadratic response cost.

\section{Optimization under a bound on the quench amplitude}
\label{appC}
In this appendix, we derive the quench vector that maximizes
the local quadratic work over the region
$B_\epsilon
=
\left\{
\Delta\bm{\lambda}:
\Vert\Delta\bm{\lambda}\Vert\leq\epsilon
\right\}.$

In the response eigenbasis introduced in Sec.~\ref{sec24}, the quench
vector is written as
$\Delta\bm{\lambda}
=
\sum_{\kappa_k>0}\xi_k\bm v_k,$
and the local quadratic work becomes
\begin{equation}
W_{\rm loc}
=
\sum_{\kappa_k>0}
\left(
b_k\xi_k-\kappa_k\xi_k^2
\right).
\label{c2}
\end{equation}


We first maximize Eq.~\eqref{c2} without imposing the amplitude
bound. Maximization with respect to each channel amplitude gives
$\xi_{k,\rm unc}
=
b_k/(2\kappa_k).$
The resulting unconstrained-optimal  quench vector is
\begin{equation}
\Delta\bm{\lambda}_{\rm unc}
=
\frac{1}{2}
\sum_{\kappa_k>0}
\frac{b_k}{\kappa_k}\bm v_k,
\label{c22}
\end{equation}
with amplitude
$\xi_{\rm unc}
=\Vert\Delta\bm{\lambda}_{\rm unc}\Vert$.

Evaluating Eq.~\eqref{c2} at $\xi_{k,\rm unc}$ gives the
maximum local quadratic work in the absence of the amplitude bound,
\begin{equation}
W_{\rm loc,unc}^{\max}
=
\frac{1}{4}
\sum_{\kappa_k>0}
\frac{b_k^2}{\kappa_k}.
\label{c33}
\end{equation}

If $\Vert\Delta\bm{\lambda}_{\rm unc}\Vert\leq\epsilon$, then
$\Delta\bm{\lambda}_{\rm unc}$ lies in $B_\epsilon$ and also
maximizes $W_{\rm loc}$ over this region.

In the following, we consider the case
$\Vert\Delta\bm{\lambda}_{\rm unc}\Vert>\epsilon$.
Since $W_{\rm loc}$ is strictly concave in the subspace spanned by
the channels with $\kappa_k>0$, its unique stationary point is
$\Delta\bm{\lambda}_{\rm unc}$. Because this point lies outside
$B_\epsilon$, the maximum of $W_{\rm loc}$ over $B_\epsilon$
cannot occur in the interior of this region and must therefore lie
on its boundary,
$\Vert\Delta\bm{\lambda}\Vert=\epsilon.$
Using the expansion
$\Delta\bm{\lambda}
=\sum_{\kappa_k>0}\xi_k\bm v_k$
and the orthonormality relation
$\bm v_k^{T}\bm v_l=\delta_{kl}$, we obtain
\begin{align}
\Vert\Delta\bm{\lambda}\Vert^2
&=
\Delta\bm{\lambda}^T\Delta\bm{\lambda}
\nonumber\\
&=
\left(
\sum_{\kappa_k>0}\xi_k\bm v_k
\right)^T
\left(
\sum_{\kappa_l>0}\xi_l\bm v_l
\right)
\nonumber\\
&=
\sum_{\substack{\kappa_k>0\\\kappa_l>0}}
\xi_k\xi_l\bm v_k^{\mathsf T}\bm v_l
\nonumber\\
&=
\sum_{\kappa_k>0}\xi_k^2.
\label{cnorm}
\end{align}

The boundary condition is therefore equivalent to
\begin{equation}
\sum_{\kappa_k>0}\xi_k^2=\epsilon^2.
\label{c3}
\end{equation}

To maximize Eq.~\eqref{c2} subject to Eq.~\eqref{c3}, we introduce
a Lagrange multiplier $\zeta>0$ and write
\begin{equation}
\mathcal{L}
=
\sum_{\kappa_k>0}
\left(
b_k\xi_k-\kappa_k\xi_k^2
\right)
-
\zeta
\left(
\sum_{\kappa_k>0}\xi_k^2-\epsilon^2
\right).
\label{c4}
\end{equation}

The stationarity condition
$\partial\mathcal{L}/\partial\xi_k=0$ gives
$\xi_{k,\rm opt}
=b_k/[2(\kappa_k+\zeta)].$ Thus, 
the corresponding constrained-optimal quench vector is therefore
\begin{equation}
\Delta\bm{\lambda}_{\rm opt}^{\epsilon}
=
\frac{1}{2}
\sum_{\kappa_k>0}
\frac{b_k}{\kappa_k+\zeta}\bm v_k.
\label{c6}
\end{equation}

The value of $\zeta$ is determined by the boundary condition
$\Vert\Delta\bm{\lambda}_{\rm opt}^{\epsilon}\Vert=\epsilon$.
Using Eqs.~\eqref{cnorm} and \eqref{c6}, this condition becomes
\begin{equation}
\sum_{\kappa_k>0}
\frac{b_k^2}{4(\kappa_k+\zeta)^2}
=
\epsilon^2.
\label{c91}
\end{equation}

At $\zeta=0$, the left-hand side of Eq.~\eqref{c91} is
$\Vert\Delta\bm{\lambda}_{\rm unc}\Vert^2
>\epsilon^2,$
whereas it approaches zero as $\zeta\to\infty$.
Since it is continuous and strictly decreasing, Eq.~\eqref{c91}
has a unique positive solution.

Evaluating Eq. (\ref{c2}) at $\xi_{k,\rm opt}$ gives the maximum local quadratic work over $B_\epsilon$,
\begin{equation}
W_{\rm loc,\epsilon}^{\max}
=
\frac{1}{4}
\sum_{\kappa_k>0}
\frac{b_k^2(\kappa_k+2\zeta)}
     {(\kappa_k+\zeta)^2}.
\label{c8}
\end{equation}

When $\Delta\bm{\lambda}_{\rm unc}$ already lies in
$B_\epsilon$, one sets $\zeta=0$, and Eqs.~\eqref{c6}
and \eqref{c8} reduce to Eqs.~(\ref{c22}) and (\ref{c33}), respectively.

For a weighted amplitude bound,
$\Delta\bm{\lambda}^T
G\Delta\bm{\lambda}
\leq\epsilon^2,$
where $G$ is symmetric and positive definite, the same
Lagrange-multiplier procedure is most conveniently carried out
in the original control basis. It gives
\begin{equation}
\Delta\bm{\lambda}_{\rm opt}^{\epsilon,G}
=
\frac{1}{2}
\left(K+\zeta G\right)^{-1}\bm b_0,
\label{c10}
\end{equation}
where $\zeta$ is chosen so that
$\left(
\Delta\bm{\lambda}_{\rm opt}^{\epsilon,G}
\right)^{T}
G\Delta\bm{\lambda}_{\rm opt}^{\epsilon,G}
=\epsilon^2$, 
when the bound is reached. Setting $G=I$ recovers the Euclidean
result in Eq.~\eqref{c6}.

\section{Asymptotic near-optimality for the exact output work}
\label{appF}
In this appendix, we show that the quench vector that maximizes
the local quadratic work over the admissible small-quench region
also yields an asymptotically near-maximum exact output work.
Let $B_\epsilon
=\left\{
\Delta\bm{\lambda}:
\Vert\Delta\bm{\lambda}\Vert\leq\epsilon
\right\}.$
Since $W_{\rm out}$ is smooth in a neighborhood of
$\Delta\bm{\lambda}=\bm 0$, Taylor's theorem implies that there
exist constants $\epsilon_0>0$ and $C>0$ such that
\begin{equation}
\left|
W_{\rm out}(\Delta\bm{\lambda})
-W_{\rm loc}(\Delta\bm{\lambda})
\right|
\leq
C\Vert\Delta\bm{\lambda}\Vert^3
\label{d1}
\end{equation}
for all $\Vert\Delta\bm{\lambda}\Vert\leq\epsilon_0$.

For any $0<\epsilon\leq\epsilon_0$, let
$\Delta\bm{\lambda}_{\rm opt}^{\epsilon}$ maximize
$W_{\rm loc}$ over $B_\epsilon$, and let
$\Delta\bm{\lambda}_{\rm ext}^{\epsilon}$ maximize
$W_{\rm out}$ over the same region. Since $B_\epsilon$ is compact
and both work functions are continuous, these maximizers exist.
Explicitly,
\begin{equation}
\Delta\bm{\lambda}_{\rm opt}^{\epsilon}
\in
\underset{\Delta\bm{\lambda}\in B_\epsilon}{\arg\max}\,
W_{\rm loc}(\Delta\bm{\lambda}),
\label{d2}
\end{equation}
and
\begin{equation}
\Delta\bm{\lambda}_{\rm ext}^{\epsilon}
\in
\underset{\Delta\bm{\lambda}\in B_\epsilon}{\arg\max}\,
W_{\rm out}(\Delta\bm{\lambda}).
\label{d3}
\end{equation}

Define the Taylor remainder by
$R(\Delta\bm{\lambda})
=W_{\rm out}(\Delta\bm{\lambda})
-W_{\rm loc}(\Delta\bm{\lambda}).$
Because $\Delta\bm{\lambda}_{\rm opt}^{\epsilon}$ maximizes
$W_{\rm loc}$ over $B_\epsilon$,
$W_{\rm loc}
\left(\Delta\bm{\lambda}_{\rm ext}^{\epsilon}
\right)-W_{\rm loc}
\left(\Delta\bm{\lambda}_{\rm opt}^{\epsilon}
\right)
\leq0.$
It therefore follows that
\begin{equation}
    \begin{aligned}
&W_{\rm out}
(\Delta\boldsymbol{\lambda}^{\epsilon}_{\rm ext})
-W_{\rm out}
\left(\Delta\boldsymbol{\lambda}^{\epsilon}_{\rm opt}
\right)
\\
&=
W_{\rm loc}
\left(
\Delta\boldsymbol{\lambda}^{\epsilon}_{\rm ext}
\right)
-W_{\rm loc}
\left(
\Delta\boldsymbol{\lambda}^{\epsilon}_{\rm opt}
\right)
\\
&\quad+
R\left(
\Delta\boldsymbol{\lambda}^{\epsilon}_{\rm ext}
\right)
-
R
\left(
\Delta\boldsymbol{\lambda}^{\epsilon}_{\rm opt}
\right)
\\
&\leq
\left|
R
\left(
\Delta\boldsymbol{\lambda}^{\epsilon}_{\rm ext}
\right)
\right|
+
\left|
R
\left(
\Delta\boldsymbol{\lambda}^{\epsilon}_{\rm opt}
\right)
\right|
\\
&\leq
2C\epsilon^3.
\end{aligned}
\end{equation}

The lower bound follows from the fact that
$\Delta\bm{\lambda}_{\rm ext}^{\epsilon}$ maximizes
$W_{\rm out}$ over $B_\epsilon$. Hence,
\begin{equation}
0
\leq
W_{\rm out}
\left(
\Delta\bm{\lambda}_{\rm ext}^{\epsilon}
\right)
-
W_{\rm out}
\left(
\Delta\bm{\lambda}_{\rm opt}^{\epsilon}
\right)
\leq
2C\epsilon^3.
\label{d5}
\end{equation}

Therefore, the exact output work attained by
$\Delta\bm{\lambda}_{\rm opt}^{\epsilon}$ differs from the maximum
exact output work over $B_\epsilon$ by at most a term of order
$\epsilon^3$.

\section{Response-cost eigenvalues for the two-level model}
\label{appB}

In this appendix, we give the detailed derivation of the radial and
transverse eigenvalues of the matrix $ K$ for the two-level working medium.

We first derive the static response matrix for the two-level working medium. Its Hamiltonian reads
\begin{equation}
  \hat H(\bm{\lambda})=\lambda_x\hat\sigma_x+\lambda_z\hat\sigma_z.  
\end{equation}

Let \(r = |\boldsymbol{\lambda}| = \sqrt{\lambda_x^2 + \lambda_z^2}\) and define the unit vector \(\boldsymbol{n} = \boldsymbol{\lambda}/r\). The Hamiltonian can then be written as \(\hat{H} = r\,\boldsymbol{n}\cdot\hat{\boldsymbol{\sigma}}\). Using the Pauli-matrix identity \((\boldsymbol{n}\cdot\hat{\boldsymbol{\sigma}})^2 = \hat{I}\), the Gibbs state takes the form
\begin{equation}
\hat\rho_\beta
=\frac{e^{-\beta r\bm n\cdot\hat{\bm\sigma}}}
{{\rm Tr}\,\left[e^{-\beta r\bm n\cdot\hat{\bm\sigma}}\right]}
=
\frac{1}{2}
\left[
\hat{I}-\tanh(\beta r)\bm n\cdot\hat{\bm\sigma}
\right].
\end{equation}

With the identity
\({\rm Tr}(\hat\sigma_\alpha\hat\sigma_\mu)=2\delta_{\alpha\mu}\), the
expectation value of the controlled operators is
\begin{equation}
    \langle\hat O_\mu\rangle_{\beta,\bm{\lambda}}
=
{\rm Tr}(\hat\rho_\beta\hat\sigma_\mu)
=
-\tanh(\beta r)n_\mu,
\qquad
\mu=x,z .
\end{equation}

The response matrix is defined by
\((D^{(\beta)})_{\mu\nu}
=\partial_{\lambda_\nu}
\langle\hat O_\mu\rangle_{\beta,\bm{\lambda}}\). Differentiating the expression above yields
\begin{align}
\left(D^{(\beta)}\right)_{\mu\nu}
&=
\frac{\partial}{\partial\lambda_\nu}
\left[
-\tanh(\beta r)n_\mu
\right]
\nonumber\\
&=
-
\left[
\frac{\partial\tanh(\beta r)}{\partial\lambda_\nu}n_\mu
+
\tanh(\beta r)
\frac{\partial n_\mu}{\partial\lambda_\nu}
\right].
\label{Dderivative}
\end{align}

We now evaluate the two partial derivatives in Eq.~\eqref{Dderivative}. Using \(r = \sqrt{\lambda_x^2+\lambda_z^2}\), we obtain \(\partial r/\partial \lambda_\nu = \lambda_\nu / r = n_\nu\). It follows that
\begin{equation}
    \frac{\partial\tanh(\beta r)}{\partial\lambda_\nu}
=
\beta\,{\rm sech}^2(\beta r)n_\nu .
\label{p1}
\end{equation}

Next, from the definition \(n_\mu = \lambda_\mu / r\), we compute
\begin{equation}
    \frac{\partial n_\mu}{\partial\lambda_\nu}
=
\frac{\delta_{\mu\nu}}{r}
-
\frac{\lambda_\mu\lambda_\nu}{r^3}
=
\frac{\delta_{\mu\nu}-n_\mu n_\nu}{r}.
\label{p2}
\end{equation}

Substituting Eqs. (\ref{p1}) and (\ref{p2}) into Eq.~\eqref{Dderivative} yields the component form of the response matrix:
\begin{align}
\left(D^{(\beta)}\right)_{\mu\nu}
&=
-
\beta\,{\rm sech}^2(\beta r)n_\mu n_\nu
-
\frac{\tanh(\beta r)}{r}
\left(
\delta_{\mu\nu}-n_\mu n_\nu
\right).
\end{align}

Equivalently, in matrix form $D^{(\beta)}
=
-\beta\,{\rm sech}^2(\beta r)\bm n\bm n^T
-
\tanh(\beta r)/r
\left(
\mathbb{I}-\bm n\bm n^T
\right).$
Evaluating this expression at the reference point
\(\bm{\lambda}_0=r_0\bm n_0\), we arrive at
\begin{equation}
D^{(\beta)}
=
-\beta\,{\rm sech}^2(\beta r_0)\bm n_0\bm n_0^T
-
\frac{\tanh(\beta r_0)}{r_0}
\left(
\mathbb{I}-\bm n_0\bm n_0^T
\right).
\label{Dtwo}
\end{equation}

We now diagonalize the response matrix and then construct the
corresponding response-cost matrix \(K\). We introduce the radial and transverse projectors
\(P_\parallel = \boldsymbol{n}_0\boldsymbol{n}_0^T,\qquad
P_\perp = \mathbb{I}- \boldsymbol{n}_0\boldsymbol{n}_0^T,\)
so that Eq.~\eqref{Dtwo} is recast as
\begin{equation}
    D^{(\beta)}
=-A_\beta P_{\parallel}-
B_\beta P_{\perp},
\end{equation}
with \(A_\beta = \beta\,\mathrm{sech}^2(\beta r_0)\) and \(B_\beta = \tanh(\beta r_0)/r_0\).
Let the radial unit vector be \(\boldsymbol{v}_\parallel = \boldsymbol{n}_0\). Using \(P_\parallel \boldsymbol{n}_0 = \boldsymbol{n}_0\) and \(P_\perp \boldsymbol{n}_0 = \boldsymbol{0}\), we compute
\begin{align}
D^{(\beta)}\bm v_{\parallel}
&=
\left(
-A_\beta P_{\parallel}
-
B_\beta P_{\perp}
\right)\bm n_0
\nonumber\\
&=
-A_\beta \bm n_0
\nonumber\\
&=
-\beta\,{\rm sech}^2(\beta r_0)\bm v_{\parallel}.
\end{align}

The radial eigenvalue of \(D^{(\beta)}\) is thus \(d_\parallel^{(\beta)} = -\beta\,\mathrm{sech}^2(\beta r_0)\).

We now take a transverse unit vector \(\boldsymbol{v}_\perp\) obeying \(\boldsymbol{v}_\perp^T \boldsymbol{n}_0 = 0\). One has \(P_\parallel \boldsymbol{v}_\perp = \boldsymbol{0}\) and \(P_\perp \boldsymbol{v}_\perp = \boldsymbol{v}_\perp\), so
\begin{align}
D^{(\beta)}\bm v_{\perp}
&=
\left(
-A_\beta P_{\parallel}
-
B_\beta P_{\perp}
\right)\bm v_{\perp}
\nonumber\\
&=
-B_\beta \bm v_{\perp}
\nonumber\\
&=
-\frac{\tanh(\beta r_0)}{r_0}
\bm v_{\perp}.
\label{app:eq:D_transverse_eigen}
\end{align}

This gives the transverse eigenvalue \(d_\perp^{(\beta)} = -\tanh(\beta r_0)/r_0\).
We define the response-cost matrix $ K
=-\left(D^{(\beta_h)}+D^{(\beta_c)}\right)/2.$
Since both \(D^{(\beta_h)}\) and \(D^{(\beta_c)}\) are evaluated at the common reference point \(\boldsymbol{\lambda}_0 = r_0\boldsymbol{n}_0\), they share identical radial and transverse eigenvectors. Applying $ K$ to the radial eigenvector yields
\begin{align}
 K\bm v_{\parallel}
&=
-\frac{1}{2}
\left(
D^{(\beta_h)}+D^{(\beta_c)}
\right)\bm v_{\parallel}
\nonumber\\
&=
-\frac{1}{2}
\left[
-\beta_h{\rm sech}^2(\beta_h r_0)
-
\beta_c{\rm sech}^2(\beta_c r_0)
\right]\bm v_{\parallel}
\nonumber\\
&=
\frac{1}{2}
\left[
\beta_h{\rm sech}^2(\beta_h r_0)
+
\beta_c{\rm sech}^2(\beta_c r_0)
\right]\bm v_{\parallel}.
\end{align}

The corresponding radial eigenvalue of $ K$ is
\begin{equation}
    \kappa_{\parallel}
=
\frac{1}{2}
\left[
\beta_h{\rm sech}^2(\beta_h r_0)
+
\beta_c{\rm sech}^2(\beta_c r_0)
\right].
\end{equation}

Likewise, acting with $ K$ on the transverse eigenvector gives
\begin{align}
K\bm v_{\perp}
&=
-\frac{1}{2}
\left(
D^{(\beta_h)}+D^{(\beta_c)}
\right)\bm v_{\perp}
\nonumber\\
&=
-\frac{1}{2}
\left[
-\frac{\tanh(\beta_h r_0)}{r_0}
-
\frac{\tanh(\beta_c r_0)}{r_0}
\right]\bm v_{\perp}
\nonumber\\
&=
\frac{1}{2r_0}
\left[
\tanh(\beta_h r_0)
+
\tanh(\beta_c r_0)
\right]\bm v_{\perp}.
\end{align}

The transverse eigenvalue is then 
\begin{equation}
    \kappa_{\perp}
=
\frac{1}{2r_0}
\left[
\tanh(\beta_h r_0)
+
\tanh(\beta_c r_0)
\right].
\end{equation}

In the radial–transverse basis, we have \(K\boldsymbol{v}_\parallel = \kappa_\parallel \boldsymbol{v}_\parallel\) and \(K\boldsymbol{v}_\perp = \kappa_\perp \boldsymbol{v}_\perp\), meaning that $K$ is  diagonal in this basis. For this two-dimensional control space, the transverse subspace is one-dimensional, so the radial–transverse set forms a complete eigenbasis for $K$.

\section{Fixed-amplitude optimization of the exact output work}
\label{appE}
In this appendix, we describe the numerical procedure for determining the quench vector that maximizes the exact output work at each fixed amplitude $\xi$.

For the XYZ spin-chain working medium, the control vector and the
corresponding quench vector are ordered as
$\boldsymbol{\lambda}=(J_x,J_y,J_z,h)^T$ and
$\Delta\boldsymbol{\lambda}
=(\Delta J_x,\Delta J_y,\Delta J_z,\Delta h)^T$, respectively.
For each value of $\xi$, the optimization is performed over all
directions of the quench vector in this four-dimensional control space.

To cover these directions, we parameterize the quench vector as
\begin{equation}
\Delta\boldsymbol{\lambda}
(\theta_1,\theta_2,\theta_3;\xi)
=
\xi
\begin{pmatrix}
\cos\theta_1\\
\sin\theta_1\cos\theta_2\\
\sin\theta_1\sin\theta_2\cos\theta_3\\
\sin\theta_1\sin\theta_2\sin\theta_3
\end{pmatrix},
\end{equation}
with $0\leq\theta_1\leq\pi$,
$0\leq\theta_2\leq\pi$, and $0\leq\theta_3<2\pi$.
This parameterization covers all quench directions at the prescribed
amplitude and satisfies
$\left|\Delta\boldsymbol{\lambda}
(\theta_1,\theta_2,\theta_3;\xi)
\right|=\xi$
by construction.

For each trial quench vector, we use the symmetric parameterization
$\boldsymbol{\lambda}_{h,c}
=
\boldsymbol{\lambda}_0
\pm
\Delta\boldsymbol{\lambda}/2$
and  evaluate the exact output work using Eq.~(\ref{wsys}) together with
\(W_{\rm out}=-W_{\rm cyc}^{\rm sys}\).
The angles that maximize the exact output work at amplitude $\xi$ are defined by
\begin{equation}
(\theta_1^\ast,\theta_2^\ast,\theta_3^\ast)
=
\arg\max_{\theta_1,\theta_2,\theta_3}
W_{\rm out}
\left[
\Delta\boldsymbol{\lambda}
(\theta_1,\theta_2,\theta_3;\xi)
\right],
\label{E2}
\end{equation}

Here the angular variables are restricted to the ranges specified above. For each fixed amplitude $\xi$, the maximization in Eq.~(\ref{E2}) was performed over the full angular domain using a differential-evolution global search. All quench vectors that maximize the exact output work satisfy the heat-engine conditions $W_{\rm out}>0$, $Q_h>0$, and $Q_c<0$ throughout the interval $0<\xi\le0.5$. The corresponding quench vector that maximizes the exact output work is
\begin{equation}
\Delta\boldsymbol{\lambda}^{\rm ext}_{\rm opt}(\xi)
=
\Delta\boldsymbol{\lambda}
(\theta_1^\ast,\theta_2^\ast,\theta_3^\ast;\xi).
\end{equation}


\begin{thebibliography}{99}
\bibitem{Sco59}
H.~E.~D. Scovil and E.~O. Schulz-DuBois,
Three-level masers as heat engines,
Phys. Rev. Lett. \textbf{2}, 262 (1959).

\bibitem{Ali79}
R.~Alicki,
The quantum open system as a model of the heat engine,
J. Phys. A: Math. Gen. \textbf{12}, L103 (1979).


\bibitem{Gev94}
E.~Geva and R.~Kosloff,
Three-level quantum amplifier as a heat engine: A study in finite-time thermodynamics,
Phys. Rev. E \textbf{49}, 3903 (1994).

\bibitem{Fel00}
T.~Feldmann and R.~Kosloff,
Performance of discrete heat engines and heat pumps in finite time,
Phys. Rev. E \textbf{61}, 4774 (2000).



\bibitem{Kos14}
R.~Kosloff and A.~Levy,
Quantum heat engines and refrigerators: Continuous devices,
Annu. Rev. Phys. Chem. \textbf{65}, 365 (2014).

\bibitem{Goo16}
J.~Goold, M.~Huber, A.~Riera, L.~del Rio, and P.~Skrzypczyk,
The role of quantum information in thermodynamics---A topical review,
J. Phys. A: Math. Theor. \textbf{49}, 143001 (2016).





\bibitem{Can24}
L.~M. Cangemi, C.~Bhadra, and A.~Levy,
Quantum engines and refrigerators,
Phys. Rep. \textbf{1087}, 1 (2024).



\bibitem{Qua07}
H.~T. Quan, Y. Liu, C.~P. Sun, and F.~Nori,
Quantum thermodynamic cycles and quantum heat engines,
Phys. Rev. E \textbf{76}, 031105 (2007).

\bibitem{Kos17}
R.~Kosloff and Y.~Rezek,
The quantum harmonic Otto cycle,
Entropy \textbf{19}, 136 (2017).

\bibitem{Aba12}
O.~Abah, J.~Ro{\ss}nagel, G.~Jacob, S.~Deffner, F.~Schmidt-Kaler, K.~Singer, and E.~Lutz,
Single-ion heat engine at maximum power,
Phys. Rev. Lett. \textbf{109}, 203006 (2012).

\bibitem{Ros16}
J.~Ro{\ss}nagel, S.~T. Dawkins, K.~N. Tolazzi, O.~Abah, E.~Lutz, F.~Schmidt-Kaler, and K.~Singer,
A single-atom heat engine,
Science \textbf{352}, 325 (2016).

\bibitem{Pet19}
J.~P.~S. Peterson, T.~B. Batalh{\~a}o, M.~Herrera, A.~M. Souza, R.~S. Sarthour, I.~S. Oliveira, and R.~M. Serra,
Experimental characterization of a spin quantum heat engine,
Phys. Rev. Lett. \textbf{123}, 240601 (2019).

\bibitem{Kla19}
J.~Klatzow, J.~N. Becker, P.~M. Ledingham, C.~Weinzetl, K.~T. Kaczmarek, D.~J. Saunders, J.~Nunn, I.~A. Walmsley, R.~Uzdin, and E.~Poem,
Experimental demonstration of quantum effects in the operation of microscopic heat engines,
Phys. Rev. Lett. \textbf{122}, 110601 (2019).

\bibitem{Bou21}
Q.~Bouton, J.~Nettersheim, S.~Burgardt, D.~Adam, E.~Lutz, and A.~Widera,
A quantum heat engine driven by atomic collisions,
Nat. Commun. \textbf{12}, 2063 (2021).

\bibitem{Zha02}
J.-W. Zhang, J.-Q. Zhang, G.-Y. Ding, J.-C. Li, J.-T. Bu, B.~Wang, L.-L. Yan, S.-L. Su, L.~Chen, F.~Nori, {\c S}.~K. {\"O}zdemir, F.~Zhou, H.~Jing, and M.~Feng,
Dynamical control of quantum heat engines using exceptional points,
Nat. Commun. \textbf{13}, 6225 (2022).

\bibitem{Blo12}
I.~Bloch, J.~Dalibard, and S.~Nascimb{\`e}ne,
Quantum simulations with ultracold quantum gases,
Nat. Phys. \textbf{8}, 267 (2012).

\bibitem{Bla12}
R.~Blatt and C.~F. Roos,
Quantum simulations with trapped ions,
Nat. Phys. \textbf{8}, 277 (2012).

\bibitem{Geo14}
I.~M. Georgescu, S.~Ashhab, and F.~Nori,
Quantum simulation,
Rev. Mod. Phys. \textbf{86}, 153 (2014).

\bibitem{Mon21}
C.~Monroe, W.~C. Campbell, L.-M. Duan, Z.-X. Gong, A.~V. Gorshkov, P.~W. Hess, R.~Islam, K.~Kim, N.~M. Linke, G.~Pagano, P.~Richerme, C.~Senko, and N.~Y. Yao,
Programmable quantum simulations of spin systems with trapped ions,
Rev. Mod. Phys. \textbf{93}, 025001 (2021).



\bibitem{Hol15}V. Holubec and A. Ryabov, Efficiency at and near maximum
power of low-dissipation heat engines, Phys. Rev. E \textbf{92}, 052125 (2015).

\bibitem{Ma24} T. J. Zhou, Y. H. Ma, and C. P. Sun, Finite-time optimization of quantum szilard heat engine, Phys. Rev. Res. 6, 043001 (2024)




\bibitem{Ye22}
Z. Ye, F. Cerisola, P. Abiuso, J. Anders, M. Perarnau-Llobet, and V. Holubec,
Optimal finite-time heat engines under constrained control,
Phys. Rev. Research \textbf{4}, 043130 (2022).



\bibitem{Cak17} 
S. Çakmak, F. Altintas, A. Gençten, and
\"{O}. E. M\"{u}stecaplıo\v{g}lu,
Irreversible
work and internal friction in a quantum Otto cycle of a single arbitrary spin, Eur. Phys. J. D \textbf{71}, 75 (2017).

\bibitem{Aba19}
O.~Abah, M.~Paternostro, and E.~Lutz,
Shortcut-to-adiabaticity Otto engine: A twist to finite-time thermodynamics,
Phys. Rev. E \textbf{99}, 022110 (2019).

\bibitem{Che19b}
J.-F.~Chen, C.-P.~Sun, and H.~Dong,
Achieve higher efficiency at maximum power with finite-time quantum Otto cycle,
Phys. Rev. E \textbf{100}, 062140 (2019).

\bibitem{Klo19}
M.~Kloc, P.~Cejnar, and G.~Schaller,
Collective performance of a finite-time quantum Otto cycle,
Phys. Rev. E \textbf{100}, 042126 (2019).



\bibitem{Lee20}
S.~Lee, M.~Ha, J.-M.~Park, and H.~Jeong,
Finite-time quantum Otto engine: Surpassing the quasistatic efficiency due to friction,
Phys. Rev. E \textbf{101}, 022127 (2020).

\bibitem{Sol20}
A.~Solfanelli, M.~Falsetti, and M.~Campisi,
Nonadiabatic single-qubit quantum Otto engine, Phys. Rev. B \textbf{101}, 054513 (2020).

\bibitem{Ins20}
A.~R.~Insinga, The quantum friction and optimal finite-time performance of the quantum Otto cycle,
Entropy \textbf{22}, 1060 (2020).

\bibitem{Sar21}
S.~Saryal and B.~K.~Agarwalla,
Bounds on fluctuations for finite-time quantum Otto cycle,
Phys. Rev. E \textbf{103}, L060103 (2021).

\bibitem{Fei22}
Z.~Fei, J.-F.~Chen, and Y.-H.~Ma,
Efficiency statistics of a quantum Otto cycle, Phys. Rev. A \textbf{105}, 022609 (2022).

\bibitem{Xia23} Y. Xiao, D. Liu, J. He, L. Zhuang, W.-M. Liu, L.-L. Yan, and J. Wang, Thermodynamics and fluctuations in finite-time quantum heat engines under reservoir squeezing,
Phys. Rev. Res. \textbf{5}, 043185 (2023).

\bibitem{McK25}
T.~McKeever, O.~Diba, and A.~Nazir,
Finite-time quantum Otto engine subject to control noise and enhancement techniques,
Phys. Rev. Res. \textbf{7}, 033172 (2025).
\bibitem{Liu26} D. H. Liu and J. H. Wang, Enhancing quantum heat engine performance via unitary-enabled exponential
speedup of thermalization, Phys. Rev. E \textbf{113}, 044121 (2026).











\bibitem{Wat26}
R.~S. Watson and K.~V. Kheruntsyan,
Enhanced performance of sudden-quench quantum Otto cycles via multi-parameter control,
Phys. Rev. A \textbf{113}, 063319 (2026).

\bibitem{Sil08}
A.~Silva, Statistics of the work done on a quantum critical system by quenching a control parameter,
Phys. Rev. Lett. \textbf{101}, 120603 (2008).

\bibitem{Dor12}
R.~Dorner, J.~Goold, C.~Cormick, M.~Paternostro, and V.~Vedral,
Emergent thermodynamics in a quenched quantum many-body system,
Phys. Rev. Lett. \textbf{109}, 160601 (2012).



\bibitem{Arr19}
E.~G. Arrais, D.~A. Wisniacki, A.~J. Roncaglia, and F.~Toscano,
Work statistics for sudden quenches in interacting quantum many-body systems,
Phys. Rev. E \textbf{100}, 052136 (2019).


\bibitem{Cal11}
P.~Calabrese, F.~H.~L. Essler, and M.~Fagotti,
Quantum quench in the transverse-field Ising chain,
Phys. Rev. Lett. \textbf{106}, 227203 (2011).

\bibitem{Wat25}
R.~S. Watson and K.~V. Kheruntsyan,
Universal principles for sudden quench quantum Otto engines,
Phys. Rev. E \textbf{112}, 034120 (2025).
\bibitem{Aba18}
O.~Abah and E.~Lutz,
Performance of shortcut-to-adiabaticity quantum engines,
Phys. Rev. E \textbf{98}, 032121 (2018).

\bibitem{Cak19}
B.~{\c C}akmak and {\"O}.~E. M{\"u}stecapl{\i}o{\u g}lu,
Spin quantum heat engines with shortcuts to adiabaticity,
Phys. Rev. E \textbf{99}, 032108 (2019).



\bibitem{Ped23}
A.~Pedram, S.~C. Kad{\i}o{\u g}lu, A.~Kabak{\c c}{\i}o{\u g}lu, and {\"O}.~E. M{\"u}stecapl{\i}o{\u g}lu,
A quantum Otto engine with shortcuts to thermalization and adiabaticity,
New J. Phys. \textbf{25}, 113014 (2023).
\bibitem{Dan21} R. Dann and R. Kosloff, Inertial theorem: Overcoming
the quantum adiabatic limit, Phys. Rev. Res. \textbf{3}, 013064 (2021).

\bibitem{Don19} J.-F. Chen, C.-P. Sun, and H. Dong, Boosting the performance of quantum Otto heat engines, Phys. Rev. E \textbf{100}, 032144 (2019).











\bibitem{Jar16}
J.~Jaramillo, M.~Beau, and A.~del Campo,
Quantum supremacy of many-particle thermal machines,
New J. Phys. \textbf{18}, 075019 (2016).

\bibitem{Che19}
Y.-Y. Chen, G.~Watanabe, Y.-C. Yu, X.-W. Guan, and A.~del Campo,
An interaction-driven many-particle quantum heat engine and its universal behavior,
npj Quantum Inf. \textbf{5}, 88 (2019).

\bibitem{Kel20}
T.~Keller, T.~Fogarty, J.~Li, and T.~Busch,
Feshbach engine in the Thomas-Fermi regime,
Phys. Rev. Res. \textbf{2}, 033335 (2020).

\bibitem{Li18}
J.~Li, T.~Fogarty, S.~Campbell, X.~Chen, and T.~Busch,
An efficient nonlinear Feshbach engine,
New J. Phys. \textbf{20}, 015005 (2018).

\bibitem{Nau24}
V.~V. Nautiyal, R.~S. Watson, and K.~V. Kheruntsyan,
A finite-time quantum Otto engine with tunnel-coupled one-dimensional Bose gases,
New J. Phys. \textbf{26}, 063033 (2024).

\bibitem{Saj25}
V.~M.~Sajitha, B.~Santra, M.~J. Davis, and L.~A. Williamson,
Quantum thermal machine regimes in the transverse-field Ising model,
Phys. Rev. A \textbf{111}, 062213 (2025).







\bibitem{Koc23}
J.~Koch, K.~Menon, E.~Cuestas, S.~Barbosa, E.~Lutz, T.~Fogarty, T.~Busch, and A.~Widera,
A quantum engine in the BEC--BCS crossover,
Nature \textbf{621}, 723 (2023).

\bibitem{Sim23}
E.~Q. Simmons, R.~Sajjad, K.~Keithley, H.~Mas, J.~L. Tanlimco, E.~Nolasco-Martinez, Y.~Bai, G.~H. Fredrickson, and D.~M. Weld,
Thermodynamic engine with a quantum degenerate working fluid,
Phys. Rev. Res. \textbf{5}, L042009 (2023).

\bibitem{Wat25a}
R.~S. Watson and K.~V. Kheruntsyan,
Quantum many-body thermal machines enabled by atom-atom correlations,
SciPost Phys. \textbf{18}, 190 (2025).




\bibitem{Wil24}
L.~A. Williamson and M.~J. Davis,
Many-body enhancement in a spin-chain quantum heat engine,
Phys. Rev. B \textbf{109}, 024310 (2024).


\bibitem{Bou23}
M.~Boubakour, T.~Fogarty, and T.~Busch,
Interaction-enhanced quantum heat engine,
Phys. Rev. Res. \textbf{5}, 013088 (2023).


\bibitem{Wan12}
J.~Wang, Z.~Wu, and J.~He,
Quantum Otto engine of a two-level atom with single-mode fields,
Phys. Rev. E \textbf{85}, 041148 (2012).

\bibitem{Meh17}
V.~Mehta and R.~S. Johal,
Quantum Otto engine with exchange coupling in the presence of level degeneracy,
Phys. Rev. E \textbf{96}, 032110 (2017).




\bibitem{Azi14}
M.~Azimi, L.~Chotorlishvili, S.~K. Mishra, T.~Vekua, W.~H{\"u}bner, and J.~Berakdar,
Quantum Otto heat engine based on a multiferroic chain working substance,
New J. Phys. \textbf{16}, 063018 (2014).

\bibitem{Alt15}
F.~Altintas and {\"O}.~E. M{\"u}stecapl{\i}o{\u g}lu, General formalism of local thermodynamics with an example:
Quantum Otto engine with a spin-$1/2$ coupled to an arbitrary spin,
Phys. Rev. E \textbf{92}, 022142 (2015).

\bibitem{Cho16}
L.~Chotorlishvili, M.~Azimi, S.~Stagraczy{\'n}ski, Z.~Toklikishvili, M.~Sch{\"u}ler, and J.~Berakdar,
Superadiabatic quantum heat engine with a multiferroic working medium,
Phys. Rev. E \textbf{94}, 032116 (2016).


\bibitem{Are24}
V.~R. Arezzo, D.~Rossini, and G.~Piccitto,
Many-body quantum heat engines based on free-fermion systems,
Phys. Rev. B \textbf{109}, 224309 (2024).

\bibitem{Har20}
A.~Hartmann, V.~Mukherjee, W.~Niedenzu, and W.~Lechner,
Many-body quantum heat engines with shortcuts to adiabaticity,
Phys. Rev. Res. \textbf{2}, 023145 (2020).

\bibitem{Sol23}
A.~Solfanelli, G.~Giachetti, M.~Campisi, S.~Ruffo, and N.~Defenu,
Quantum heat engine with long-range advantages,
New J. Phys. \textbf{25}, 033030 (2023).

\bibitem{Sun18} J. Chen, H. Dong, and C.-P. Sun, Bose-Fermi duality in a quantum Otto heat engine with trapped repulsive bosons, Phys. Rev. E \textbf{98}, 062119 (2018).


\bibitem{Siv12}
D.~A.~Sivak and G.~E.~Crooks,
Thermodynamic Metrics and Optimal Paths,
Phys. Rev. Lett. \textbf{108}, 190602 (2012).

\bibitem{Zul12}
P.~R.~Zulkowski, D.~A.~Sivak, G.~E.~Crooks, and M.~R.~DeWeese,
Geometry of thermodynamic control,
Phys. Rev. E \textbf{86}, 041148 (2012).

\bibitem{Def13}
S.~Deffner and E.~Lutz,
Thermodynamic length for far-from-equilibrium quantum systems,
Phys. Rev. E \textbf{87}, 022143 (2013).

\bibitem{Bra20}
K.~Brandner and K.~Saito,
Thermodynamic Geometry of Microscopic Heat Engines,
Phys. Rev. Lett. \textbf{124}, 040602 (2020).







\bibitem{Per24}
F.~T.~B.~P\'erez and J.~M.~Matera,
Quantum covariance scalar products and efficient estimation of maximum-entropy projections,
Phys. Rev. A \textbf{109}, 022401 (2024).

\bibitem{Lac25}
A.~M.~Lacerda, L.~P.~Bettmann, and J.~Goold,
Information geometry of transitions between quantum nonequilibrium steady states,
Phys. Rev. E \textbf{112}, L022101 (2025).

\bibitem{Bet25}
L.~P.~Bettmann and J.~Goold,
Information geometry approach to quantum stochastic thermodynamics,
Phys. Rev. E \textbf{111}, 014133 (2025).


\bibitem{Hag05} R.~Hagemans, J.-S.~Caux, and U.~L\"ow,
Gapped Heisenberg spin chains in a field,
Phys. Rev. B \textbf{71}, 014437 (2005).

\bibitem{Shi19} N. Shiraishi, Proof of the absence of local conserved quantities in the XYZ chain with a magnetic field, EPL 128, 17002 (2019).

 \bibitem{Zha22} X. Zhang, A. Kl\"umper, and V. Popkov, Invariant subspaces and elliptic spin-helix states in the integrable open
spin-1/2 XYZ chain, Phys. Rev. B 106, 075406 (2022).

\bibitem{Hub57} R. Kubo, Statistical-mechanical theory of irreversible processes. I. General theory and simple applications to magnetic and conduction problems, J. Phys. Soc. Jpn. \textbf{12}, 570 (1957).

\bibitem{Wil67} R. M. Wilcox, Exponential operators and parameter differentiation in quantum physics, J. Math. Phys. 8, 962 (1967).

\end{thebibliography}
\end{document}